\documentclass[journal]{IEEEtran}

\usepackage{xcolor,soul,framed} 

\colorlet{shadecolor}{yellow}

\usepackage[pdftex]{graphicx}
\graphicspath{{../pdf/}{../jpeg/}}
\DeclareGraphicsExtensions{.pdf,.jpeg,.png}

\usepackage[cmex10]{amsmath}

\usepackage{array}
\usepackage{mdwmath}
\usepackage{mdwtab}
\usepackage{eqparbox}
\usepackage{url}
\usepackage{setspace}
\usepackage{fancyvrb}
\usepackage{listings}
\usepackage{verbatimbox}
\usepackage{xcolor}

\definecolor{codegreen}{rgb}{0,0.6,0}
\definecolor{codegray}{rgb}{0.5,0.5,0.5}
\definecolor{codepurple}{rgb}{0.58,0,0.82}
\definecolor{backcolour}{rgb}{0.95,0.95,0.92}

\lstdefinestyle{mystyle}{
    backgroundcolor=\color{backcolour},   
    commentstyle=\color{codegreen},
    keywordstyle=\color{magenta},
    numberstyle=\tiny\color{codegray},
    stringstyle=\color{codepurple},
    basicstyle=\ttfamily\footnotesize,
    breakatwhitespace=false,         
    breaklines=true,                 
    captionpos=b,                    
    keepspaces=true,                 
    numbers=left,                    
    numbersep=5pt,                  
    showspaces=false,                
    showstringspaces=false,
    showtabs=false,                  
    tabsize=2
}

\usepackage{algorithm,algpseudocode}
\usepackage{amsmath}
\usepackage{amsfonts}
\usepackage{amssymb}

\begin{document}
\bstctlcite{IEEEexample:BSTcontrol}
    \title{Threat Modelling in Internet of Things (IoT) Environment Using Dynamic Attack Graphs}
  \author{Marwa Salayma,\textit{ Member, IEEE}\\\textit{Department of Computing, Imperial College London\\} {London, United Kingdom}

  \thanks{This work was supported by PETRAS National Centre of Excellence for IoT Systems Cybersecurity (PETRAS 2), Grant number is EP/S035362/1.}
  }

\maketitle

\begin{abstract}
This work presents a threat modelling approach to represent changes to the attack paths through an Internet of Things (IoT) environment when the environment changes dynamically, i.e., when new devices are added or removed from the system or when whole sub-systems join or leave. The proposed approach investigates the propagation of threats using attack graphs. However, traditional attack graph approaches have been applied in static environments that do not continuously change such as the Enterprise networks, leading to static and usually very large attack graphs. In contrast, IoT environments are often characterised by dynamic change and interconnections; different topologies for different systems may interconnect with each other dynamically and outside the operator control. Such new interconnections lead to changes in the reachability amongst devices according to which their corresponding attack graphs change. This requires dynamic topology and attack graphs for threat and risk analysis. In this paper, a threat modelling approach is developed that copes with dynamic system changes that may occur in IoT environments and enables identifying attack paths whilst allowing for system dynamics.  Dynamic topology and attack graphs were developed that are able to cope with the changes in the IoT environment rapidly by maintaining their associated graphs. 
To motivate the work and illustrate the proposed approach,  an example scenario based on healthcare systems is introduced. The proposed approach is implemented using a Graph Database Management Tool (GDBM)- Neo4j- which is a popular tool for mapping, visualising and querying the graphs of highly connected data, and is efficient in providing a rapid threat modelling mechanism, which makes it suitable for capturing security changes in the dynamic IoT environment. 

\end{abstract}


\IEEEpeerreviewmaketitle


\section{Introduction}\label{intro}

\IEEEPARstart{W}{e} live in a time where most aspects in life are turning into digital, relying on highly connected devices that connect with each other and to the Internet leading to the so-called Internet of Things or IoT, which critical systems rely on, for example in healthcare \cite{SORRI2022121623, saravanan2022iot}. It is  assumed that control over the installation, integration and usage of IoT devices lies with the user. In contrast to the traditional computers or cloud servers securely hosted in offices or contained within secure physical locations, IoT devices have dynamic aspects and are deployed in a physical environment that can be subjected to direct connections and both physical and cyber attacks. In many cases, systems are composed of several devices that join and leave a network dynamically or that may be intermittently connected. Such devices are often mobile, either because they are mobile themselves e.g., autonomous vehicles or drones, or because they are instrumenting objects that are physically mobile e.g., body sensor networks for healthcare. 
\\
The diverse and advanced capabilities deployed by IoT devices, including the short range communication protocols, enable IoT devices not only to interconnect with each other but with other Enterprise networks existing in the same environment, allowing a device or a whole network to join or leave other networks on-the-fly in an ad-hoc manner \cite{agmon2019deployment}. When devices move location, they become exposed to new threats and cyber attacks that were not considered before, even if they had previously been thought to be secure. Understanding how to maintain the operation of a system when it has been partially compromised is therefore of critical importance. 
\\
The key to achieve this is to find ways to enable a dynamic system to cope and respond to changes, which is a significant challenge and has not been addressed in the literature. One way to model threats and propagation of attacks in an IoT environment is to use the attack graphs. An attack graph is an attack modelling technique that depicts the attack paths followed by an attacker across the network to compromise a target taking into account the vulnerabilities present in each device/host in the network and other reachable devices/hosts. Attack graphs shows us the multistage of attacks but also help us decide on appropriate countermeasures to mitigate the effect of the system compromise \cite{1021806}. Legacy approaches of attack graph modelling use static algorithms that do not adapt to changes in network devices and hosts configuration, their underlying topology and/or security objectives \cite{noel2010measuring}. Calculating the attack graph when the topology of the system changes dynamically requires calculating the transitive closure i.e., reachability over the graph on the topology to determine the reachable nodes. Although a number of algorithms exist to handle such changes in large graphs \cite{veloso2014reachability, jin2012scarab}, 
the problem does not have known solutions for dynamic graphs. Having gained privileges on a particular device an attack can either exploit another vulnerability to elevate privileges on the same device or exploit a vulnerability on any other reachable host. It is therefore necessary to calculate the reachability to all other hosts from a particular one. 
\\
An interactive modelling and analysis of an attack graph is crucial to cope with changes in the network topology (we use topology as a synonym for a system) e.g., when a new device/host is added or removed from the system, or when the the configurations of host/devices change along with the network security objectives. The need of interactive modelling of rich and highly interconnected topologies, which are prone to continuous updates and are typically associated with large attack graphs, led us to consider the use of \textit{graph database}. Graph database models information by storing the data in nodes and connect them trough edges in a graph like, which is -as opposed to the relational database that is strict in terms of data modelling- more suitable to model, update and query large, rich and highly connected data. Hence, graph database is an efficient methodology to model both the highly dynamic and interconnected topologies such the as IoT networks and their large associated attack graphs \cite{barik2014graph, barik2016network}. 
\\
Neo4j is a Graph Database Management (GDBM) system that provides optimised and native graph storage and processing capabilities in which relationships attached to a node directly connect that node to other related nodes allowing an easy way to traverse the graph and query graph paths and nodes. Neo4j adopts Cypher; a Neo4j built in graph database query language that is specifically designed for graph-based reasoning e.g through querying graph paths traversing the graph without the need of using an index for joining data used in SQL relational database query language, which is slower than the graph database in many orders of magnitude \cite{chen2016comparison}. This is because the relational database requires expensive Cartesian products (join operations) with complexity of $O(n^d)$ for traversing a graph of $n$ nodes and of depth $d$ \cite{noel2014big}. In Neo4j, however, one can traverse the graph through the direct edges that connect a node to other nodes, which means graph traversal complexity will depend only on the size of the resulted sub-graph and is not related to the total graph \cite{yuan2020attack}. 
Using Cypher, one can query a graph database implemented in Neo4j to look for data matching a specific pattern. Neo4j has rich support for querying graph paths allowing easy security analysis of attack graphs. 
\\
Besides modelling network topology graph for any network exist in the environment using Neo4j, in this work we have developed seven algorithms (implemented as Cypher queries) to investigate the propagation of attacks when the IoT environment changes using a dynamic topology and attack graphs. These algorithms are developed to:
(i) Generate the reachability graph automatically from a network modelled topology graph
(ii) Generate the attack graph from the reachability graph
(iii) Merge two topology graphs when an existing network joins another one, which can be extended for multiple networks join each other
(iv) Update the reachability graph only for the updated parts of the network
(v) Merge two attack graphs when an existing network joins another one
(vi) De-merge the merged topology graphs when a network leaves, and
(vii) De-merge the merged attack graphs when a network leaves.
\\
This paper is structured as follows: Section \ref{rw} discusses the few works we found in literature related to the proposed approach to model threats in a dynamic IoT environment. We explain our proposed use case scenario drawn from healthcare systems in Section \ref{usecase}, followed by identifying and modelling our graphs, i.e., the  network topology, reachability and attack graphs and their implementation using Neo4j in Section \ref{GM}. The proposed approach for maintaining the dynamic graphs through the merging process is presented in Section \ref{MergingGraphs}; de-merging is presented in Section \ref{DeMergingGraphs}. We illustrate the application to our use-case in Section \ref{runningAlgo} by running the queries for different scenarios related to the presented Use Case, while Section \ref{metrics} shows how attack graph based metrics towards analysing the risk of compromise in the dynamic IoT environment can be implemented. Finally, we present our conclusions in Section \ref{conclusion}. 

\section{Related Work} \label{rw}
\subsection{Attack Graphs}
Different ways of representing attack graphs have been adopted in the literature. Attack graphs were first defined as a state-based approach where each node in the graph represents the state of the system \cite{barik2016attack, swiler1998graph}. This approach was shown not to scale and is particularly unsuitable in IoT environments as every change to the network topology e.g., a device being added requires to change \textbf{all} the states in the model. 
\\
In contrast we adopt the so-called \textit{logical attack graph representation} where nodes in the graph represent pre- and post-conditions and exploits that can be performed by the attacker leveraging known vulnerabilities (it is in its strict sense a bi-partite graph) \cite{jajodia2009topological, jajodia2005topological}. This formulation is also compatible with the application of Bayesian techniques as described in \cite{munoz2017exact} and also with a number of tools for attack-graph generation such as Mulval \cite{ou2006scalable}. Pre- and Post- conditions in this case correspond to expressions over the privileges required to exploit the vulnerability and privileges obtained after the vulnerability exploit respectively. In their review paper, Barik et. al \cite{barik2016attack} discuss the attack graph generation and analysis techniques proposed in literature. 

\subsection{Graph Modelling Using Neo4j}
Despite its usefulness in modelling and querying highly connected and scalable data, very few work have considered using Graph Database Management (GDBM) systems to store and retrieve data related to network system and/or attacks propagation. However, information related to the network topology physical and logical components, the services installed on those components, the vulnerabilities of these services, the conditions to exploit the vulnerabilities along with the propagation of the attack are all data that can be stored and represented using Graph Database systems one of which is the Neo4j.
\\
An early work that uses Neo4j to model attack propagation in Enterprise network topologies was presented in \cite{barik2014graph}. The work adopts exploit dependency graph representation and assumes two types of nodes; entity nodes that represent the main component of an Enterprise network such as hosts, services, and their vulnerabilities, and fact nodes that represent interactions among entity nodes. However,  
the modelling of the network graph is troublesome due to the numerous types of fact nodes and edges between them, and many of the edges and nodes are not used or needed in the graph generation. This is because the work does not harness a feature of edges properties supported by Neo4j, which can help reduce the number of nodes and edges necessary to model the graph. The work assumes static reachability between hosts, and between, software services, whilst many details related to attack graph generation e.g., vulnerabilities, pre and post conditions are omitted. As a result, the generated attack graphs do not completely capture the full semantics of the network topology needed to analyse the attack paths which led to unrealistic graph modelling and query results. 
\\
The work in \cite{barik2016network} presents an attempt to improve \cite{barik2014graph} by introducing constraints that network topologies must enforce to represent the network access controls between machines necessary for the generation of the attack graphs. The work proposes a Graph Constraint Language (GCON) as an extension over the standard property Graph Model, the model that Neo4j is built upon. The work adopts the exploit dependency graph representation. However, the constrains mechanisms are not clear especially when it comes to the firewall rules queries and their impact on the reachability between hosts and their software services (firewall rules seem to allow traffic in both directions). As a result, the attack path was not a one-way directed path from the initial node to the target node, which makes it difficult to query and visualize the possible attack paths. Although information associated with vulnerabilities such as pre and post conditions are mentioned, they are not used in the attack graph generation queries and the work does not explicitly model the accessibility information between hosts in the attack graph generation. As a result, the graph generation process is complex especially when the network scales. How the firewall rules queries impact the reachability between hosts, and how the reachability is calculated and queried are not clear in this work, which is similar to the work in \cite{barik2014graph}.
\\
Although they did not use Neo4j itself to generate the attack graph, Noel et al. in \cite{noel2014big} used Neo4j Cypher queries to analyse an attack graph generated using the TVA/Cauldron tool, a tool designed by George Mason University for generating and analyzing attack graphs leveraging hosts vulnerability scans and firewall configurations, according to which it determines the reachability between machines in different domains. The work proposes a shared environment model with data ingested from other sources, such as the MongoDB database, and Apache Spark for network events besides sources for network flows, IDS alerts, anti-virus logs and so on. This shared environment is used as a database input to create a graph in Neo4j, which is then visualized, queried and analyzed for the security state of a network and how attackers can incrementally proceed their attack. \\
In their following work in \cite{Noel2016Chapter4}, the same authors developed an extended version of the shared environment -the CyGraph- to improve network security.  CyGraph leverages existing tools and data sources to build a knowledge graph. Similar to the work in  \cite{noel2014big}, CyGraph uses the TVA/Cauldron tool to generate the attack graph for a network that considers firewall rules, host configurations, and vulnerabilities. CyGraph uses Neo4j to store the graph data (nodes, relationships, and properties) as its backend database. Graph pattern-matching queries are either expressed in Neo4j Cypher query language or using the domain-specific CyGraph Query Language (CyQL), complied by CyGraph to native Cypher. 
\\
However, both works in \cite{noel2014big} and in \cite{Noel2016Chapter4} need to import data from many sources to create the agnostic model, and some of those sources are commercial tools such as the Cauldron tool, which is used to generate the attack graphs, whilst Neo4j itself can be used to generate the attack graph associated with the network topology if it is provided by means to consider the actual firewall rules and routing paths in the network . CyGraph is geared towards the enterprise environments and do not consider IoT environments and their dynamic nature. Getting data from various data sources can be troublesome if the data related to network topologies is dynamic and continuously change, as this needs to update the data in all those input resources when the environment changes, making the proposed shared environment inconvenient to model threats in the dynamic IoT environment. 
\\
The work in \cite{yuan2020attack} uses Neo4j to store information related to network machines along with their vulnerabilities. The work uses simple Cypher queries to investigate the directed attack paths initiated by exploiting a vulnerability in an initial node towards exploiting a vulnerability in the target machine. The attack graph is similar to the one in \cite{barik2014graph, barik2016network} and is generated over the topology graph. However, the way firewall rules are implemented by adding a node for each accessible port and adding edges between the nodes that can access each other is not practical especially if the network size changes or firewall configuration changes. 
\\
In all the studies discussed above, the graphs and queries implemented to generate or analyse the topology and the attack graphs consider static networks and do not account for what happens when the network environment changes, i.e., new nodes added or removed from the network or the the network configurations and firewall rules change. Additionally, all the presented efforts that use Neo4j to generate the attack graph do so by creating an attack graph over the network graph itself ending in one graph, which makes it difficult to analyse and visualise both the topology and its associated attack graph or query them. This also makes it difficult to consider that multiple network graphs exist in the environment along with their corresponding attack graphs. A separate attack graphs and network topology graphs are required for this purpose and for better visualisation, investigation, query and analysis. 
\\
Most importantly, although few works considered traffic configuration polices such as the firewall rules in their implemented static network topology, none of the work presented above shows how such packet filtering policies impact the reachability between hosts and how the attack graph is affected accordingly. How reachability between devices and hosts is calculated or how it changes when the environment changes is a major issue to consider when generating attack graph in order to get sound query results.

\subsection{Reachability Graph}
In essence, reachable nodes in graph means there is a directed path between them, and this is typically computed by taking the transitive closure (the reachability matrix) of the graph computed e.g., through the Floyd-Warshall algorithm, which states that there is a path between any two nodes in a graph if and only if there is an edge between the two nodes or there is a path between the two nodes going through any number of hops (nodes), between the two nodes in question \cite{weisstein2008floyd}. 
\\
The work in \cite{1498492} provided static modelling and representation for the reachability between routers in a TCP/IP network as a graph where the graph nodes represent the set of routers, and the graph edges represent the connectivity between the routers. The work provided an attempt to compute reachability in a TCP/IP network that includes packet filters implemented in the routers to control the traffic between the routers. \\
The work describes the reachablity between two routers as a subset of packets that the network will carry between the two routers which can be computed using a set of union and intersection operations to reduce complex operations of computing the transitive closure. 
The work differentiate between instantaneous reachability in a network at a single instant in time, the upper bound reachability referring to the largest set of packets the network will ever deliver between two points, and the lower bound reachability referring to largest set of packets the network will always deliver between two point. The work then provided approximations to the Reachability bounds. 
\\
In this work we are using Neo4j to query the reachability between devices/machines in our proposed topology graph by considering packet filtering rules implemented in routers. In contrast to \cite{1498492}, our reachability queries can be considered instantaneous, i.e., assuming that the forwarding state is known at a specific instant, which requires knowing the configuration state of each router. However, not only our reachability query can run with a low time complexity, but also can run only for the updated parts of the network whenever the network updated or configuration changes by harnessing the Neo4j graph database features as will be explained in Section \ref{reachability}.

\section{Example on our Use-case Drawn From Health Care Systems} \label{usecase}
\begin{figure}
  \begin{center}
  \includegraphics[width=\linewidth]{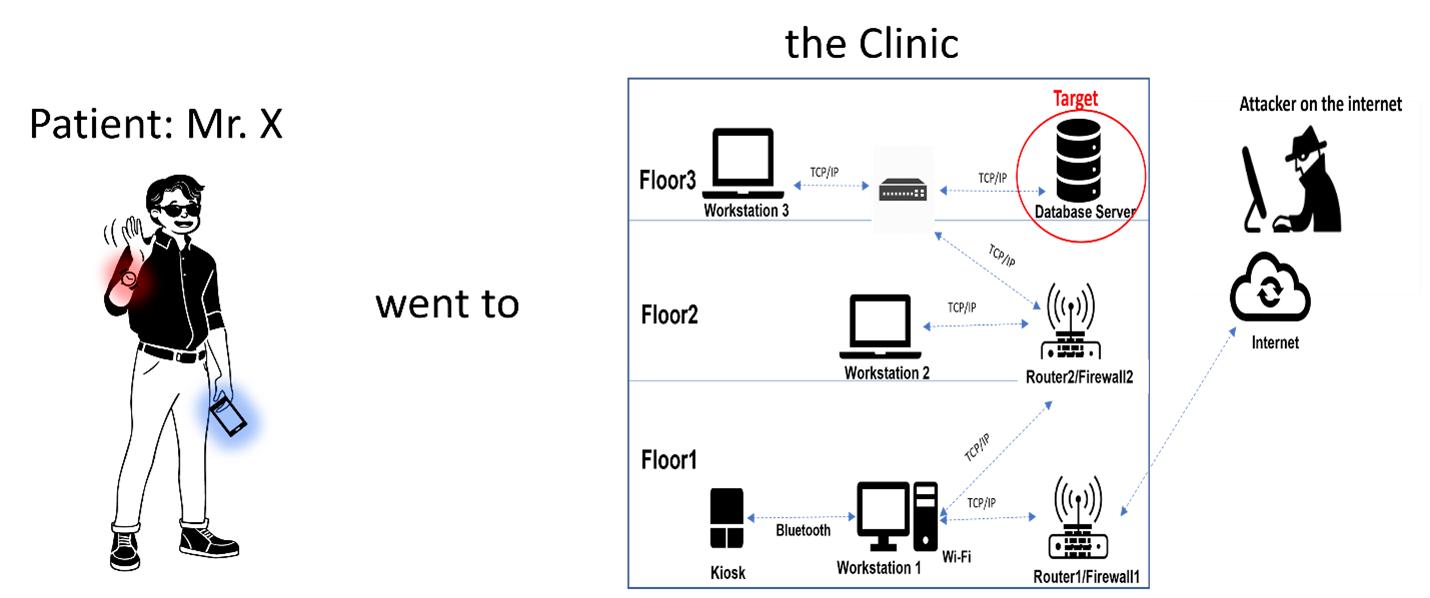}
  \caption{Example on our use-case drawn from healthcare systems (Patient and clinic).}\label{topologies}
  \end{center}
\end{figure}

As an example, we consider the scenario of a patient wearing a body sensor network (BSN) including sensors and a gateway (typically a mobile phone) visiting a clinic. This allows to reason about the patient’s devices connecting to the clinic and thus acquiring reachability across the clinic’s system to other services and devices. \\
For example, health apps on the patient’s device may need to connect to the services inside the clinic to upload monitoring data, download prescriptions for diagnostic tests or for medication. Re-calibration of the sensors on the patient may also be triggered by the clinic. The important point to model is that the devices and services inside the clinic can become reachable form the patient’s devices and vice versa thus creating new possible attack paths. By considering that the patient’s body sensor network comprises several devices and that the mobile phone acts as a routing gateway, this allows us to also consider the body sensor network as a “system” and thus to investigate what happens to the attack paths when systems become interconnected in a “systems-of-systems” approach. \\
We have considered in detail the topologies of the two networks for the patient and for the clinic as shown in Fig.~\ref{topologies}. For the clinic we have considered different points of access into the network, access to databases and, to other network servers. We have also considered the internal segmentation of the clinic’s network in sub-networks, traffic filtering, and specific vulnerabilities on the devices drawn from NVD \cite{21841}. Thus, we can model the attack paths that from the patient’s devices may lead to the database using attack graphs \cite{noel2010measuring}. We have also considered that the clinic has a connection to the Internet to consider the propagation of external attack paths across the combined systems. \\
Similarly, we have considered vulnerabilities in the devices part of the patient’s body sensor network to consider attack paths from clinic's devices across the two systems that may lead to a target device on the patient (Smart Watch). \\
It is important to note that in contrast to the method adopted in literature of generating an attack graph over the topology graph, we consider that the attack-graph for the two systems i.e., the patient’s BSN and the clinic are computed independently and independent from their associated topology graphs, thus allowing us to consider the merging and de-merging the attack graphs when topologies inter-connect and dis-connect respectively, besides allowing us analyse the security characteristics of networks in the environment in more efficient and easier manner.  
\\
Two types of attacks can be conducted in our patient-clinic use case scenario: (i)	External attack (attacker is on the internet) attacks clinic systems remotely, and (ii) internal attack which targets Bluetooth enabled devices and it is initiated internally while the attacker in close vicinity to any Bluetooth enabled device, after which the attack can proceed remotely using the internet. \\
In this work,  we are investigating the propagation of threats across both networks, and to reason about that we are considering different attackers with different targets: (1) one attack target is located in the  Clinic Topology that is the Database Server, in essence to get a root privilege over the Database sever, (2) the other attack target is located in the Patient Topology, that is the Smart Watch, in essence to get a user privilege on the patient Smart Watch. 

\section{Graphs definitions and modelling}\label{GM}
\subsection{Network Topology} \label{NT}
In our network topology we have end devices such as IoT devices or workstations, and other network devices such as routers and switches. Each router can be have different interfaces allowing it to attach to different subnets. The routers are connected with each other allowing communication between different routers associated with the different subnets. For example, in the clinic all routers will be connected to Router1 (the aggregator) which connects the Clinic to the Internet, Router 1 also has an interface that connects it to Subnet 1. We have also Router 2 that connects Subnet 1, Subnet 2 and Subnet 3 together and Router 2 is connected to Router 1. Other routers can be created in the same way and direct edges can be created between the routers and any end device if there is a direct physical connection between them. 
\\
We consider that all routers can implement packet filtering rules that can deny access by default, and only allow traffic flows that are explicitly allowed. End Devices which located in the same subnet are logically connected through the router. Hence, there is no direct edge between them. End devices that are not equipped with IP addresses can still communicate with each other e.g., through direct wireless links such as Bluetooth if they are in communication range with each other. In such cases we create a direct physical connection between the end devices to reflect this direct connectivity. 
\\
This topology design leads to having either direct connections between end devices and routers (or switches) or direct connections between end devices. All connection links between devices, can also have additional attributes such as type of protocol that the connection allows (e.g., TCP, UDP, Blutooth, ZigBee etc). Without loss of generality, in our use-case we have only considered communication via TCP connections or direct communication via Bluetooth. 
\\
With this topology design and configuration, two assumptions for the possible routes populated in the routers' routing table hold:
\begin{itemize}
    \item Each router implements a routing table only for the routes associated with the networks to which the router is directly attached to (directly connected). As a result, all the end devices whether they are in the same subnet, or in an adjacent subnet connected to the same router can communicate with each other’s, in the latter case if and only if there is a firewall rule that allows the reachability between those two end devices, and this firewall rule can be implemented in the router that connects the different subnets directly.
    \item Each router in its routing table accounts for every single network in the network topology, this includes routes for the networks to which the router is directly attached to (directly connected), as well as routs for the indirectly connected networks that can be either populated statically by the network administrator or shared dynamically between the adjacent routers through one of the networking protocols \cite{Buchanan1999, 1498492}. For example, in the clinic we can assume that each router knows about the rout for its directly connected subnets and the routes for indirectly connected subnets that are connected to the adjacent router and so on. As a result, whether the end devices in an adjacent subnets connected to the same router or nonadjacent subnets connected through multi hop routes can communicate with each other’s, if and only if there is a firewall rule that allows the reachability between those two end devices, and firewall rules are implemented in any router 
    \footnote{The routers share their the routs dynamically between each other’s using routing protocols, or statically (manually) by the network administrator leading to any end device in any subnet can reach any other end device in the clinic even if it is not in the same subnet. However, reachability between devices can be limited by denying access between devices by default and using firewall rules to explicitly state which devices can communicate with each others}. 
\end{itemize}
In this work we adopt the second the assumption, i.e., each router in its routing table accounts for every single network in the network topology through static or dynamic routing. As mentioned in Section \ref{usecase}, we have considered vulnerabilities in the end devices in both the clinic and the patient. \\
In the clinic floor 1/subnet 1, Workstation 1 machine uses Windows OS and runs a vulnerable version of a web browser, which has a vulnerability \textit{CVE-2017-6753} that allows an unauthenticated remote attacker to execute an arbitrary code and gain user privileges on that machine. Workstation 1 is also equipped with a Bluetooth adapter, which has the \textit{CVE-2017-8628} vulnerability in the Microsoft's implementation of the Bluetooth stack allowing the attacker to obtain access to higher level services and profiles and eventually obtaining overall control. 
\\
A Kiosk end device is also located in the clinic on floor 1; we assume that it communicates and is managed directly by Workstation 1 through Bluetooth. The Kiosk runs Linux OS and similar to Workstation 2 has Linux kernel stack overflow vulnerability \textit{CVE-2017-1000251} in its Bluetooth adapter. Workstation 2 in clinic floor 2/subnet 2 uses Linux OS and provides an FTP service which has a vulnerability \textit{CVE-2021-41635}, allowing an attacker to abuse the machine configurations and obtain access to the entire machine. Workstation 2 is also equipped with Bluetooth adapter, which has \textit{CVE-2017-1000251} vulnerability, that provides an attacker with a full and reliable kernel-level exploit for any Bluetooth enabled device running Linux. 
\\
Workstation 3 in clinic floor 3/subnet 3 also runs Linux and provides an SSH service which has a vulnerability \textit{CVE-2022-30318}, allowing an attacker to obtain overall control.  The Database Server, also located in floor 3/subnet 3, runs Linux with kernel v.2.6 with MySQL RDBMS v.5 which has \textit{CVE-2009-2446} vulnerability that enables an attacker to gain user privileges on the Database Server.For the purpose of the example we consider this as the end goal of an external attacker accessing from the internet. 
\\
The Firewall implemented in Router 1 allows only HTTP traffic from the Internet to devices/machines in subnet 1, i.e., Workstation 1 and blocks all other traffic. The Firewall implemented in Router 2 allows FTP and SSH traffic to Workstation 2 and to Workstation 3 respectively, as well as access to the Database Server from Workstation 2 and Workstation 3 and blocks all other traffic. Those firewall rules for Firewall 1 and Firewall 2 are listed in Table \ref{table:1} and Table \ref{table:2} respectively. 
\\
For simplicity, we refer to the devices by their names rather than their IP addresses. Such topology example along with its physical components, packet filtering configurations, software vulnerability exploits along with their associated pre and post conditions which we will explain in detail in the following sections, are expected  to provide an external attacker whose machine is on the internet with two paths to access the target, i.e., the Database Server, either by exploiting the vulnerability in Workstation 1 form where the attacker can lunch an attack on Workstation 2, which eventually leads the attacker to the Database Server, or the attacker can exploit the vulnerability in Workstation 3 from Workstation 1 which leads the attacker to exploit a vulnerability on the Database Server.   

\begin{table}[ht]
\centering
\caption{Firewall 1 rule set}\label{table:1}
\resizebox{\columnwidth}{!}{\begin{tabular}{|c|c|c|c|c|c|c|}
  \hline
  Rule& Source& Destination& srcPort & dstPort & Protocol & Action \\ [0.5ex] 
  \hline
  Rule1& Any& Subnet 1 & Any & Any & TCP & Allows \\ 
  \hline
\end{tabular}}
\end{table}

\begin{table}[ht]
\caption{Firewall 2 rule set}\label{table:2}
\centering
\resizebox{\columnwidth}{!}{\begin{tabular}{|c|c|c|c|c|c|c|}
  \hline
  Rule& Source& Destination& srcPort & dstPort & Protocol & Action \\ [0.5ex] 
  \hline
  Rule1 & Subnet 1 & Subnet 2 & Any & Any & TCP & Allows \\ 
  \hline
  Rule2 & Subnet 1 & Workstation 3 & Any & Any & TCP & Allows \\ 
  \hline
  Rule3 & Subnet 2 & Subnet 3 & Any & Any & TCP & Allows \\ 
  \hline
\end{tabular}}
\end{table}
Our patient is wearing a Smart Watch, which can connect to the patient Smart Phone and any other Bluetooth enabled device in its vicinity using a Bluetooth connection. The Smart Watch deploys sensors that sense critical physiological information such as the patient heart rate making it a target for another attacker who wants to  re-calibrate the sensors on the patient Smart Watch and manipulate the sensed data. The Smart Watch runs a Bluetooth stack in the Linux Kernel (BlueZ),which has \textit{CVE-2017-1000251}; a stack overflow vulnerability, that provides an attacker with a full and reliable kernel-level exploit. 
\\
The patient also carries a Smart Phone which has an IP address allowing it to create a TCP/IP connection with any router around it and has also a Bluetooth connection which allows it to connect with other Bluetooth devices in its vicinity. The patient's Smart Phone runs Android OS which has a Bluetooth Android information Leak \textit{CVE-2017-0785} vulnerability, allowing the attacker to access the whole phone filesystem, gain full control of a device, and use the victim’s Bluetooth interface to attack other devices in its proximity. 
\\
In the following section we show how we can model the network topologies of both the patient and the clinic.

\subsubsection{Network Topology Graph $G_N$} \label{NTG}
\begin{figure}
  \begin{center}
  \includegraphics[width=\linewidth]{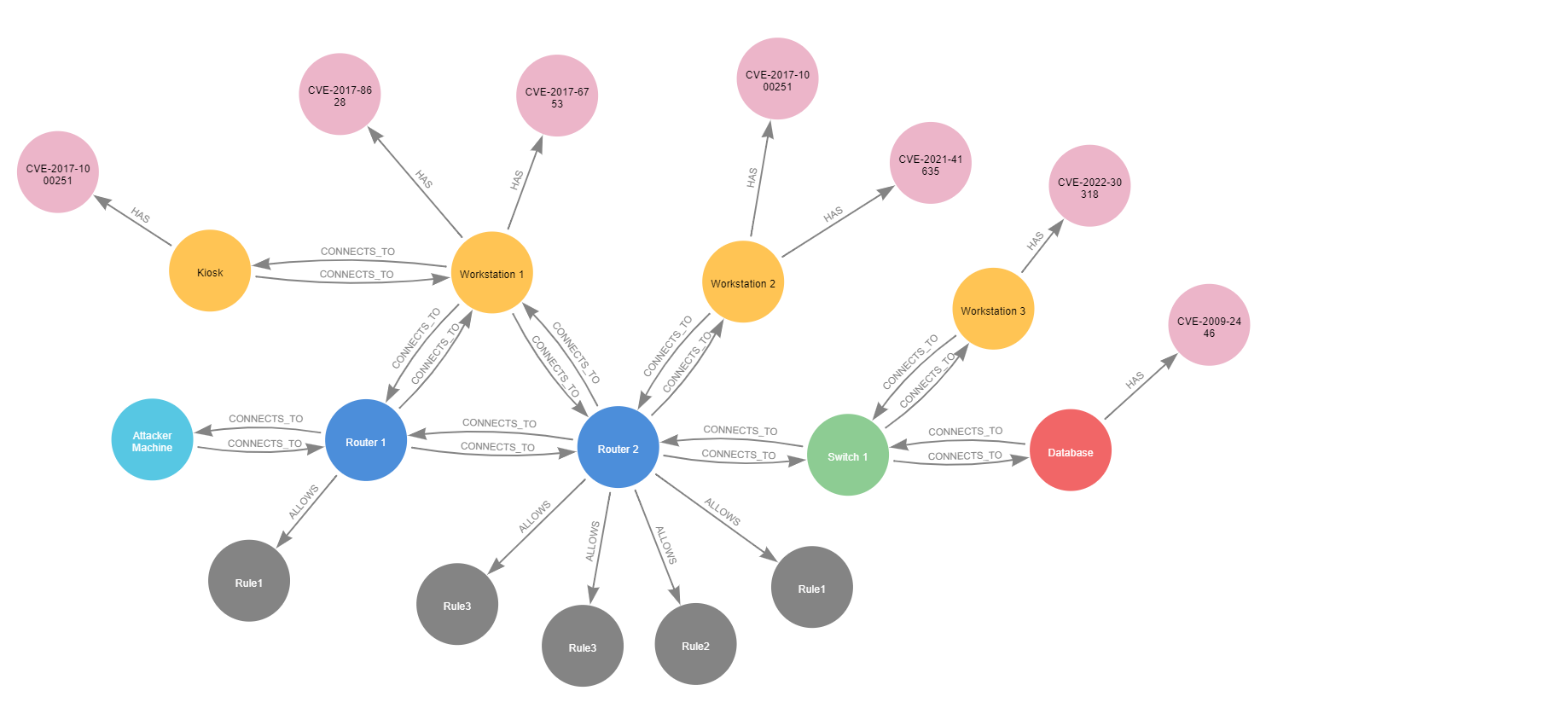}
  \caption{Clinic topology graph.}\label{clinic topology graph}
  \end{center}
\end{figure}

\begin{figure}
  \begin{center}
  \includegraphics[width=1in]{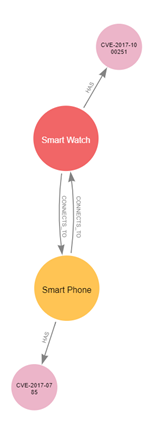}
  \caption{Patient topology graph.}\label{patient topology graph}
  \end{center}
\end{figure}

We can model the network topology of any network using a graph: $G_N=\{V, E, A, C\}$, where $V$ is the set of all the nodes in the graph, $E$ is the set of all the edges in the graph, $A$ is a set of attributes associed with nodes and edges, and $C$ is a set of categories in which nodes and edges can be grouped. We discuss each of these graph elements in detail below.
\begin{itemize}
    \item Graph nodes $V=\{V_d, V_v,V_r\}$, where $V_d=\{V_e, V_n\}$ represents set of all the devices in the network, which can be end devices $V_e$ such as IoT devices or workstations, or any other network devices $V_n$ such as routers and switches. $V_v$ is the set of vulnerabilities that may exist in devices' software services.
    $V_r$ is a set of firewall rules implemented in the routers. 
    \item Graph edges $E =\{E_d, E_v, E_f\}$ where $E_d$ is the set of all physical point to point connectivity links. This could be a direct connection between an end device and a router or switch $(v_e , v_n)$, a direct connection between a router and a switch $(v_{n_i}, v_{n_j})$, or could be a point-to-point short range connection such as Bluetooth connection between two end devices $(v_{e_i}, v_{e_j})$. $E_v$ are edges that associate end devices with their respective vulnerabilities, hence, $\forall v_e \in V_e $ if $\exists v_v \in V_v$, associated with $v_e$ then $\exists (v_e , v_v) \in E_v$. $E_f$ are the edges that associate routers with the firewall rules deployed on them, hence $\forall v_n \in V_n$ , if $\exists v_r \in V_r$ associated with $v_n$, then $\exists (v_n , v_r) \in E_f$. 
    \item Attributes $A =\{A_v, A_e\}$ represent properties of the graph nodes and edges, where $A_v = \{A_d, A_{vul}, A_r\}$ is a set of attributes that annotate the graph nodes representing the network devices, devices vulnerabilities and firewall rules respectively. $A_e$ on the other hand is a set of attributes that annotates the graph edges. For example, in our clinic topology graph, $A_d= \{name, subnet, floor, accessibility, privilege\}$. Apart from the accessibility attribute, each of the attributes has a single value. An accessibility list $a_e$ is a \textbf{set} of features that allow an end device access other devices, e.g.,  Workstation 1 has an IP address and has software services that allows access other machines, such as the FTP and the SSH servers. Note that although on the same device usually there is at least one privilege and typically there would be multiple user privileges (e.g., for all the different users or for several accounts with superuser privileges ), for simplicity in this work we assume that on each device there is only one privilege either a user or a superuser privilege, hence the privilege attribute has a single value.  Generally we say  $\forall v_d \in V_d$ $\exists a_d \in A_d$. Another example, $A_{vul}= \{preConditions, postConditions\}$ represents lists of pre-conditions required to exploit vulnerabilities and post-conditions result from successful vulnerabilities' exploits. Generally we say $\forall v_v \in V_v$ $\exists a_{vul} \in A_{vul}$. On the other hand, the set of attributes that annotates the graph edges $A_e= \{via\}$, where \textit{via} attribute adds additional information about the connection between two devices, e.g., specifying the protocol used such as TCP/IP or Bluetooth. Attribute values including the lists are stored as strings. 
    \item Categories $C =\{C_v, C_e\}$, represent the sets of labels/types that help us filter the query results, where $C_v$ is a set of labels that can be used to filter our query results for the graph nodes, e.g., by providing a name for the network topology graph to which the device belongs, or by providing the type of the device whether its an end device or a router or a switch and so on. $C_e$ is used to filter query results for the edges, e.g., it may indicate whether the edge represents a point to point connection between two devices $(v_{e_i}, v_{e_j})$ by setting the type of the edge as $CONNECTS\_TO$, or refers to  an edge between an end device that has a vulnerability and its vulnerability node $(v_e, v_v)$ by setting the type of the edge as $HAS$, or an edge of type $ALLOWS$ between a router and a firewall rule $(v_n, v_r)$ implemented in the router. Categories provide a form of identification for a group of nodes that belong to the same category. Instead of querying each node in the set of nodes by its name, if the set of those nodes belong to the same category, one can retrieve the nodes all at once by querying their category. Categories provide a very helpful way to identify the set of devices that belong to a certain topology (we use topology as a synonym for a system) amongst other network topologies available in the environment, which will enable us to retrieve all the devices in that topology. This also means that nodes in one set may or may not belong to the same category, e.g., a node $v_{e_i} \in V_e$  and $v_{e_i}\in C_1$ another node $v_{e_j} \in V_e$ but $v_{e_j} \in C_2$. Moreover, each node may belong to more than one category, for example, a device can belong to EndDevice category and that end device may also belong to the ClinicTopology category. 
\end{itemize}
To create a node in the network we first need to specify its categories. For example, to create a node representing a device, we need to specify to which network topology it belongs, the type of the device, i.e., whether it’s an end device or a network device, and to which subnet it belongs. Besides, we need to specify its attributes i.e., its name, which floor its located in and the accessibility features that allows it access other devices in the network, e.g., whether the device has an IP address and what other software services it supports such as HTTP, FTP, SSH, MYSQL and so on. \\
The topology graphs for the systems in our use case scenario which refer to the clinic and the patient are represented in Fig.~\ref{clinic topology graph} and Fig.~\ref{patient topology graph} respectively. 

\subsubsection{Neo4j Implementation for the Topology Graph $G_N$}  \label{NTGI}
The following are examples on how we are using Cypher queries to create nodes and edges in a graph to model the network topology of the clinic depicted Fig.~\ref{clinic topology graph}. \\
\textbf{Creating topology devices:} in the following Cypher query, we are referring to the Clinic Network Topology Graph $G_{N_C}$ by using the category ClinicTopology, we will be adding everything depicted in Fig.~\ref{clinic topology graph} to this category starting from Workstation 1, for which we will be setting a set of attributes. For example Workstation 1 belongs to subnet 1 and it is located in floor 1.  
\begin{Verbatim}[showspaces=false,fontsize=\small]

CREATE(n:ClinicTopology{name: ‘Workstation 1', 
subnet: ‘subnet 1', floor: ‘floor 1'})

\end{Verbatim}
We can further specify that Workstation 1 is an end device, by adding the appropriate category to Workstation 1 as follows:

\begin{Verbatim}[showspaces=false,fontsize=\small]

MATCH(n:ClinicTopology) WHERE 
n.name= ‘Workstation 1’ SET n:EndDevice 

\end{Verbatim}
In this way we are adding the node $n$ which refers to Workstation1 to the category/label $c_v= EndDevice$, which also means that we are adding Workstation 1 to the set of end devices $V_e$ in the set of devices $V_d$ in $G_{N_C}$. We may then add accessibility features as another property to Workstation 1, such as the software services that allow Workstation 1 to connect with other devices in the networks, for example, besides having the basic network protocol IP, Workstation 1 can launch FTP connection, SSH connection, Bluetooth and so on, this can be done by providing Workstation 1 with a list of all the features it deploys, as follows \footnote{For the simplicity we are considering IP which is a basic network protocol as one of accessibility feature besides the FTP and SSH, which are services that run on top of the IP that allow the device to access other devices software services, given the fact that some IoT devices doe not have IP address to access other devices software services, so those devices will not include IP in their accessibility features.}:

\begin{Verbatim}[showspaces=false,fontsize=\small]

MATCH(n:ClinicTopology:EndDevice) WHERE 
n.name =‘Workstation 1’ SET 
n.accessibility =[‘IP', ‘FTP', ‘SSH', ‘Bluetooth']

\end{Verbatim}
\textbf{Representing vulnerabilities in the graph:} Workstation 1 is found to have two vulnerabilities. We can create a node representing one of those vulnerabilities as follows:
\begin{Verbatim}[showspaces=false,fontsize=\small]

CREATE(v:ClinicTopology:Vulnerability {name:
‘CVE-2017-8628'})

\end{Verbatim}
In the above query we are adding the node $v$ that we created to the set of vulnerabilities $V_v$ that exist in the Workstation 1 software services by crating a new category called Vulnerability, and we set the name attribute of this vulnerability to \textit{CVE-2017-8628}. To exploit this vulnerability, a set of pre-conditions need to be satisfied, those conditions can be specified through the attribute preConditions associated with the vulnerability node. For example:
\begin{Verbatim}[showspaces=false,fontsize=\small]

MATCH(v:ClinicTopology:Vulnerability {name:
‘CVE-2017-8628}’}) SET 
v.preConditions =[‘User’,‘HTTP’] 

\end{Verbatim}
By setting the attribute preConditions for the vulnerability node $v$, we represent that in order to exploit the vulnerability $v$, the attacker needs to be a user on the device from where they are launching the attack and that device should support the HTTP protocol, i.e., the attacker  device should have HTTP protocol in its accessibility list. Successful exploit of the vulnerability can lead to post-conditions and those can also be specified through the vulnerability attribute postConditions. For examples:
\begin{Verbatim}[showspaces=false,fontsize=\small]

MATCH(v:ClinicTopology:Vulnerability {name:
‘CVE-2017-8628'}) SET v.postConditions =[‘User’]

\end{Verbatim}
By setting the attribute postConditions for the vulnerability $v$, we state that after successfully exploiting the vulnerability $v$, the attacker will acquire the privileges of a user on the device that has vulnerability $v$. Then in order to associate this vulnerability with Workstation 1,  we can create an edge of type (category) ‘HAS’, which also means we are adding a new edge to the set $E_v$, as follows:   
\begin{Verbatim}[showspaces=false,fontsize=\small]

MATCH(n:ClinicTopology:EndDevice)
MATCH(v:Vulnerability)
WHERE n.name =‘Workstation1’ AND v.name =
‘CVE-2017-8628’
CREATE(n)-[:HAS]->(v)

\end{Verbatim}
If we want to add a router to the topology graph, we can create category/label called Router, to which we can add all routers that exist in the topology, which also means that we are adding a router with name Router 1 to the set of network devices $V_n$ in the set of devices $V_d$. 
\begin{Verbatim}[showspaces=false,fontsize=\small]

CREATE(r:ClinicTopology:Router{name:‘Router 1’}

\end{Verbatim}
\textbf{Creating firewall rules:} our firewall rules are implemented in the routers and control communication between devices in different subnets and with the internet, and each router has its own firewall rules. Therefore, in order to write queries to state the firewall rules implemented in each router, we first need to retrieve each router individually besides retrieving all devices in all subnets using the categories associated with them, as follows:
\begin{Verbatim}[showspaces=false,fontsize=\small]

MATCH(i:ClinicTopology:Internet) 
MATCH(s1:ClinicTopology:Subnet1)
MATCH(s2:ClinicTopology:Subnet2)
MATCH(s3:ClinicTopology:Subnet3)
MATCH(r1:ClinicTopology:Router{name:‘Router 1'})
MATCH(r2:ClinicTopology:Router{name:‘Router 2'})

\end{Verbatim}
We then can write the firewall rules listed in Table \ref{table:1} and Table \ref{table:2} using Cypher queries, e.g., 
\begin{Verbatim}[showspaces=false,fontsize=\small]

MERGE(f1:ClinicTopology:Firewall{name:‘Rule1',
source: i.name, destination: s1.name, srcPort: 
‘any', dstPort: ‘any', protocol:‘TCP'})

\end{Verbatim}
In the above query, we are making each rule in the rules set in a router firewall as a node belong to the category Firewall with attributes corresponding to each item in the firewall tuples presented in Table \ref{table:1} and Table \ref{table:2} in Section \ref{NT}\footnote{MERGE clause works like CREATE but only create a node as specified in the query if that node does not exist already.}. Rule 1 in the rule set of Firewall 1 states that any host in the internet can communicate with any host in Subnet 1. Another way to do it, is by creating rules associated with each end device in Subnet 1 one by one but in this case each rule requires a separate cypher query. 
\\
In this way, we are creating a node of variable $f_1$ with category called Firewall that will include all the firewalls rules we have in the clinic topology and we are adding $f_1$ to the set of firewall rules $V_r$ that are implemented in Router 1. Note that because there is only Workstation 1 in Subnet 1, one node representing the firewall rule will be generated, and if there are other end devices in Subnet 1, more than one firewall rule will be generated all at once. The name of the firewall rule is Rule 1 and its associated between devices located in the internet and devices located in Subnet 1 in the clinic topology. As this rule allows devices on the internet to reach devices in clinic Subnet 1, we create an edge of type ALLOWS that associates this firewall rule with the firewall implemented in Router 1, as follows: 
\begin{Verbatim}[showspaces=false,fontsize=\small]

MERGE(r1)-[:ALLOWS]->(f1)

\end{Verbatim}
Thus, we are presenting the action ALLOWS in each rule as an edge between the router that implements the firewall rule set, which also means we are adding a new edge to the set $E_r$. We can do the same for the rest of the firewall rules sets and create directional edges of category ALLOWS between the routers and and their rules. \\
Although our network settings states that all traffic is denied by default, we still can also implement less strict traffic restrictions by providing a mixture of both allows and denies rules. Similar to creating ALLOWS edges, we can easily add DENIES edges to our firewall rules queries. For example, we can explicitly state that the firewall implemented in Router 2 denies access between end devices in Subnet 1 and the Database in Subnet 3, by adding the following cypher query to the firewall rule set: 

\begin{Verbatim}[showspaces=false,fontsize=\small]

MERGE(f2:ClinicTopology:Firewall{name:‘Rule2',
source: s1.name, destination: ‘Database', 
srcPort: ‘any', dstPort: ‘any', protocol:‘TCP'})
MERGE(r2)-[:DENIES]->(f2) 
    
\end{Verbatim} 
\textbf{Creating connection links (edges):} we can see from the network topology in Fig.~\ref{clinic topology graph} that Workstation 1 has a direct physical connection with Router 1, in the topology graph we can represent this connection by creating bidirectional edges with category/type CONNECTS\_TO between Workstation 1 and Router 1, as follows:
\begin{Verbatim}[showspaces=false,fontsize=\small]

MATCH(n:ClinicTopology:EndDevice)
MATCH(m:ClinicTopology:Router)
WHERE n.name =‘Workstation 1’ AND 
m.name =‘Router 1’
CREATE(n)-[:CONNECTS_TO]->(m)
CREATE(n)<-[:CONNECTS_TO]-(m)

\end{Verbatim} 
In this way we are adding connectivity edges in both directions to the set of edges $E_d$ that groups the physical point to point links between both the Workstation 1 and Router 1. We can provide an additional information related to these links by specifying an attribute that states the protocol used for this connection: 
\begin{Verbatim}[showspaces=false,fontsize=\small]

MATCH(n:ClinicTopology:EndDevice{name:
‘Workstation1’})-[r]-(m:ClinicTopology:Router
{name:‘Router 1’}) SET r.via = ‘TCP’

\end{Verbatim}
The above query enables different types of CONNECTS\_TO edges. For example, if the devices communicate through other protocols, such as the UDP, we can write the same query but setting r.via = ‘UDP'. Because typically several different protocols can be available over a connection, we can depict different types of traffic communication protocols by using different category/types of connection edges which eliminates the need of using the attribute \emph{via}. \\
For example, we can write the following Cypher query to create edges that represent connection between two devices that exchange traffic through TCP protocol: 
\begin{Verbatim}[showspaces=false,fontsize=\small]

MATCH(n:ClinicTopology:EndDevice)
MATCH(m:ClinicTopology:Router) 
WHERE n.name= ‘Workstation 1’ AND 
m.name= ‘Router 1’
CREATE(n)-[:CONNECTS_VIA_TCP]->(m) 
CREATE(n)<-[:CONNECTS_VIA_TCP]-(m)

\end{Verbatim}
And we can write the following cypher query to create edges that represent connection between two devices that exchange traffic through ‘UDP' protocol:
\begin{Verbatim}[showspaces=false,fontsize=\small]

MATCH(n:ClinicTopology:EndDevice)
MATCH (m:ClinicTopology:Router) WHERE 
n.name =‘Workstation 2’ AND m.name =‘Router 1’
CREATE(n)-[:CONNECTS_VIA_UDP]->(m) 
CREATE(n)<-[:CONNECTS_VIA_UDP]-(m)

\end{Verbatim}
Without loss of generality, in our use-case patient-clinic scenario we are assuming one protocol for exchanging traffic between IP enabled devices, that is the TCP protocol. 

\subsection{Reachability between End-Devices} \label{reachability}
Not only firewall rules can determine reachability between devices, but reachability  is also controlled according to  the construction of the routing table on each router. A router uses specific network address to determine which out-link to use to reach a subnet, conceptually, this is called a route. Hence, routing can be considered as a kind of a dynamic traffic filtering, which is the reason why we are jointly considering traffic packet filtering and routing  together to determine reachability between end devices in our network topology graphs. \\ 
As it stated in Section \ref{NT}, a router can infer about a route in multiple ways: (i) a local routes that allow the router to reach all directly-connected subnets, (ii) through static routes specified manually to map one or more router interfaces to a destination subnet, (iii) routes shared dynamically between routers using a routing protocol, e.g., OSPF \cite{1498492}. Accordingly, two devices can reach each other if one or more of the following options apply:
\begin{itemize}
    \item The two devices are in the same subnet.
    \item The two devices connect with each other directly e.g through a short range communication protocol such as the Bluetooth.
    \item The two devices indirectly connected through one or several routers (if they have IP addresses) if and only if the firewall rules allow, in this case the devices can be in different subnets directly connected to a router or in different subnets connected to different routers that share the routes statically or manually. 
\end{itemize}
After modelling our network topology using a graph, we can generate the reachability graph from the topology graph. This is described in the following section.
\\
\subsubsection{Reachability Graph $G_R$} \label{RG}
Let $G_R= \{V_e, E_r\}$ represents our reachability graph, where as is explained in Section \ref{NTG}, $V_e$ represents the set of end devices such as IoT devices or workstations, and  $E_r$ describes the edges between end devices that are reachable from each other. A device can either have an IP address, which allows it to connect directly or indirectly with other devices in the network if the firewall rules allow, or it may not have an IP address if only direct communication is supported. However, not having an IP address will not hinder the end device from connecting and reaching directly other devices in the network using its unique MAC address or any other identifier depending on the communication technology a device adopts. We can generate the reachability graph from topology graph by considering that two end devices can reach each others if: i) the two end devices are in the same subnet, ii) the two end devices have a direct point-to-point connection or iii) the two end devices have an indirect connection. Hence, we can say that two end devices $v_{e_i}$, $v_{e_j} \in V_e$ can reach other $iff$ any of the following conditions apply: (i) $v_{e_i}.subnet = v_{e_j}.subnet$ OR (ii) $\exists (v_{e_i}, v_{e_j}) \in E_d$ OR (iii) $\exists$ an indirect connection between $v_{e_i}$ and $v_{e_j}$ such that $\forall e_d \in E_d$ between $v_{e_i}$ and $v_{e_j}$ $e_d.via = 'TCP' $ AND there exists a router with a firewall rule, i.e., $\exists v_n \in V_n$ AND $v_r \in V_r$ AND  $\exists (v_n, v_r) \in E_f$ such that $v_r.source = v_{e_i}.name$ and $v_r.destination = v_{e_j}.name$. Note we have only one category of edges $e_f \in E_f$  between a router and a firewall rule that is ALLOWS. 
\\
\subsubsection{Neo4j Implementation for the Reachability Graph ($G_R$)}
\begin{figure}
  \begin{center}
  \includegraphics[width=3in]{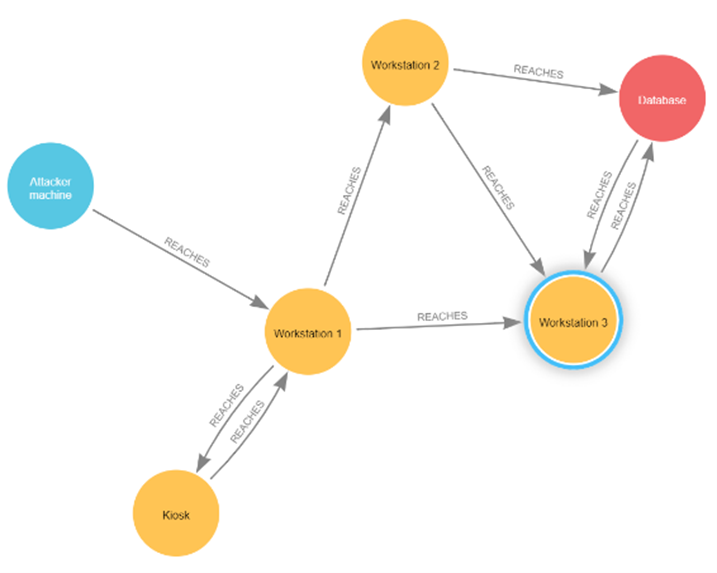}
  \caption{Clinic reachability graph.}\label{clinic reachability graph}
  \end{center}
\end{figure}

\begin{figure}
  \begin{center}
  \includegraphics[width=1.4in]{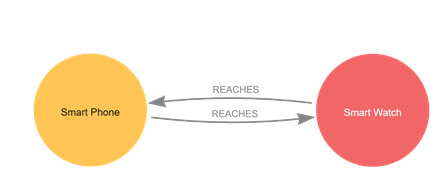}
  \caption{Patient reachability graph.}\label{patient reachability graph}
  \end{center}
\end{figure}

Our reachability graph includes only the end devices from the topology graph, and involves only one new category/type of edges $E_r$, that is REACHES. We say that nodes in a graph are reachable if there is a path between them, and this is typically computed by taking the transitive closure (the reachability matrix) of the graph e.g., through the Floyd-Warshall algorithm, which states that there is a path between any two nodes in a graph if and only if there is a direct edge between them or there is a path between the two nodes going through any number of hops (nodes) between the two nodes \cite{weisstein2008floyd}. By using the Cypher query language in Neo4j, we can compute the reachability and generate the reachability graph between devices. 
\\
Calculating the reachability from our network topology graph, requires looking at the edges between end devices in our topology graph as we assume our reachability graph includes only the end devices. Nevertheless, as it is explained in Section \ref{RG}, one reason of reachability between end devices is that the two end devices are located in the same subnet, but as long as in our topology modelling we do not create and edge between nodes in one subnet in our network graph, we need only to check the subnet property of the two end devices. 
\\
Note that although the firewall rules presented in Table\ref{table:1} and in Table\ref{table:2} show explicitly state what devices can reach each other, we do not explicitly specify that devices in same subnet can reach each other as they do so by default, this is why we need to add this case as an individual case for reachability between end devices. Moreover, after running different experiments we realised that considering nodes in the same subnet reach each other as an indvidual reachability case helps in reducing the number of database hits \footnote{A database hit is the abstract unit for retrieving or updating  data to and from the storage engine requested by Neo4j operators} as well the query response time as compared to removing this case and adding the firewall rule to show explicit access between end devices in same subnet instead.
\\
Calculating the reachability for the rest of the cases presented above is very similar to the way we calculate the transitive closure following Floyd-Warshall algorithm. As stated in Section \ref{RG}, having a direct edge between two end devices means those devices are reachable through point to point connections; a case that can happen between IoT devices. Another case is that a path between two end devices through one hop can exist if the two end devices are connected directly to a router, i.e., they are located in two different subnets to which a router is directly connected or connected indirectly through any number of routers (hops).
\\
In Neo4j, a path between two nodes can be expressed using the asterisk (*). For example, (n)-[*2]-\>(m) denotes exactly 2 relationships and one hop in between nodes n and m. [*]  without specifying bounds describes a path of at least 1 hop but of any positive length, i.e., any/infinite number of hops, allowing us to query the existence of any indirect path between any two nodes. However, due to the firewall rules that constrain traffic flows, the existence of a path does not necessarily mean that those devices are reachable. It is also important to check that the firewall rules allow the respective traffic flows. 
\\
Thus, finding the indirect path through the * should be accompanied with checking that a firewall exist in any router (by default end devices are not allowed to communicate with each other unless explicitly stated otherwise by the firewall rules).  Having said that, the reachability graph can be generated from our network topology graph using the following Cypher query: 
\begin{Verbatim}[showspaces=false,fontsize=\small]

MATCH(n:EndDevice)  
MATCH(m:EndDevice)
WHERE n.name <> m.name AND (n.subnet=m.subnet  
OR EXISTS((n)-[:CONNECTS_TO]->(m))  
OR EXISTS((n)-[:CONNECTS_TO*{via:‘TCP'}]->(m)) 
AND EXISTS((:Router)-[:ALLOWS]->(:Firewall 
{source:n.name, destination:m.name}))) 
MERGE(n)-[:REACHES]->(m)

\end{Verbatim}
The query above finds all the end devices in all the networks, and for any two different end devices n and m, create an edge of category/type REACHES between them, if any following conditions apply:
1.	the two devices are in the same subnet. 
2.	the two devices are connected directly e.g., via a point-to-point link.
3.	The two end devices are connected to each other indirectly (e.g., can communicate through TCP/IP), and there is firewall rule implemented in any router that allows communication between n and m. \\
The * operator means that starting from node n and ending in node m the graph is traversed to look for edges with the required protocol attributes, in this case via =‘TCP'. According to the reachability query results we can add a new edge REACHES to the set $E_r$. 
\\
We can also add a query to check the existence of a path associated with UDP protocol between any two devices, i.e., we can add the following query to the above reachability query:
\begin{Verbatim}[showspaces=false,fontsize=\small]

OR EXISTS((n)-[:CONNECTS_TO*{via:‘UDP'}]->(m))

\end{Verbatim}
However, in case we created different types of CONNECTS\_TO edges, for example CONNECTS\_VIA\_TCP, CONNECTS\_VIA\_UDP, we can write the following query that allows us to traverse the graph for all types of edges associated with the traffic protocols we are interested in all at once. For example we can add the following: 
\begin{Verbatim}[showspaces=false,fontsize=\small]

OR EXISTS(
(n)-[:CONNECTS_VIA_TCP|CONNECTS_VIA_UDP*]->(m)) 

\end{Verbatim}
The above reachability query allows us to generate the reachability graph between any two end devices across all the network topology graphs that exist in the environment all at once, as we are not filtering the query to find end devices from a specific network graph or with a specific category (we can filter as we discussed in Section 1.1).
\\
However, this may slow down getting the query outcome when the graph in the environment are large. For example, by running the above reachability query for all the end devices exist in all the networks used in our use-case, requires 272 ms with a total of 201529 database hits in the Cypher query run time. The vast majority of those hits, lies in finding the path comprised of CONNECTS\_TO edges. 
\\
We can limit the query by specifying which topology graph we are interested in. For example, we can query the clinic topology graph alone by providing the label ClinicTopology in the query as follows:
\begin{Verbatim}[showspaces=false,fontsize=\small] 

MATCH(n:ClinicTopology:EndDevice)
MATCH(m:ClinicTopology:EndDevice)

\end{Verbatim}
By focussing on the clinic topology alone, the query takes 191 ms, with total 49598 database hits. We conducted an additional experiment by gradually adding five more routers to the clinic topology, each with one subnet and one machine connected with each other over a chain of TCP connections. 
\\
We ran the reachability query for the clinic and checked the the query performance, which is revealed to be almost the same in terms of the query running time and the number of database hits as compared to original clinic topology. By specifying the end devices in the patient topology alone, we achieved the query result in 300 ms, with total 87 database hits, because the patient topology does not involve TCP connections. 
\\
Although the pattern (n)-[*]-\>(m) allows us query about paths between two nodes connected indirectly through several routers, the performance of a query involving this pattern could be reduced as the network size scales. Not only we can can harness filtering by specifying categories, but we can also filter using appropriate attributes. 
\\
For instance, one does not need to query the whole network graph to compute the reachability whenever the network changes especially if the network administrators know where the update/change in the network occurs, e.g in a specific floor in the clinic. To find the reachability between end devices located in floor 1 in the clinic topology, one can modify  the first two lines of the reachability query by specifying the floor attribute:
\begin{Verbatim}[showspaces=false,fontsize=\small] 

MATCH(n:ClinicTopology:EndDevice{
floor:'floor1'}) 
MATCH(m:ClinicTopology:EndDevice{
floor:'floor1'}) 

\end{Verbatim}
By specifying that we want to find reachability for end devices located in floor 1, we achieved the results in 10 ms, with only 204 database hits in total. We can also find the reachability between end devices located in different floors, e.g., floor 1 and floor 3, by specifying the floors we are interested in as follows:
\begin{Verbatim}[showspaces=false,fontsize=\small] 

MATCH(n:ClinicTopology:EndDevice{
floor:'floor1'}) 
MATCH(m:ClinicTopology:EndDevice{
floor:'floor3'}) 

\end{Verbatim}
Our experiment to find the reachability between end devices located in floor 1 and floor 3 took only 6 ms to run and 369 database hits in total. This is a significant improvement in the reachability query performance as opposed to running the query for the whole network. 
\\
In fact, the reachability query is the only query that traverses the graph searching for TCP connections as well as checking the firewall rules, making the processing time of the other queries used in this work negligible when compared to the reachability query, as the rest of all other queries require only checking directly connected nodes and none try to find out paths with indirect nodes.
\\
If one wants to consider a mixture of ALLOWS and DENIES firewall rules, then one can run a reachability query that considers a firewall implemented in a router with a firewall rule that denies access between two end devices. To do so, the reachability query presented above can be easily amended by adding one more condition that checks whether there exist a firewall connected to a router with an edge of category type DENIES, as follows:

\begin{Verbatim}[showspaces=false,fontsize=\small]

MATCH(n:EndDevice)  
MATCH(m:EndDevice)
WHERE n.name <> m.name AND (n.subnet=m.subnet  
OR EXISTS((n)-[:CONNECTS_TO]->(m))  
OR EXISTS((n)-[:CONNECTS_TO*{via:‘TCP'}]->(m)) 
AND EXISTS((:Router)-[:ALLOWS]->(:Firewall 
{source:n.name, destination:m.name})) AND NOT 
EXISTS((:Router)-[:DENIES]->(:Firewall 
{source:n.name, destination:m.name})))
MERGE(n)-[:REACHES]->(m)

\end{Verbatim}
The reachability graphs that correspond to the clinic and the patient topology graphs are represented in Fig.~\ref{clinic reachability graph} and Fig.~\ref{patient reachability graph}, which show the output that matches the reachability described through the firewall rules listed in Table \ref{table:1} and Table \ref{table:2}.  

\subsection {Attack propagation}
How attack propagates in a certain topology can be modelled through an attack graph, which is a representation of all possible attack paths an attacker may follow to compromise critical resources. It uses the knowledge base of known vulnerabilities and attack techniques on a network and then finds the different sequences of exploits and attack paths starting from attacker’s initial state, leading to compromise of critical network assets \cite{agmon2019deployment}. 
\\
To generate the attack graph for a system, in this instance the clinic and the patient, it is necessary to first consider the topology of the respective networks, which are represented in Fig.~\ref{clinic topology graph} and Fig.~\ref{patient topology graph} respectively, in order to generate their associated reachability graphs. 
\\
Next, from the reachability graph we can generate the attack graph, by considering the vulnerabilities present in each of the reachable devices, the pre-conditions necessary to exploit each vulnerability and the privileges (i.e., post-conditions) obtained after the exploitation of vulnerabilities.  In short, in order to generate an attack we need to know the following information:

\begin{itemize}
    \item The topology of the network.
    \item Devices in the topology.
    \item The vulnerabilities that may exist on those devices.
    \item The preconditions that need to be satisfied in order to successfully exploit those vulnerabilities, and whatever post-conditions result as an outcome of the successful vulnerability exploit. 
    \item Traffic limiting configurations such as firewall rules that control the reachability between the devices in the topology.
\end{itemize}

\subsubsection{Attack Graph ($G_A$)}
\begin{figure}
  \includegraphics[width=3in]{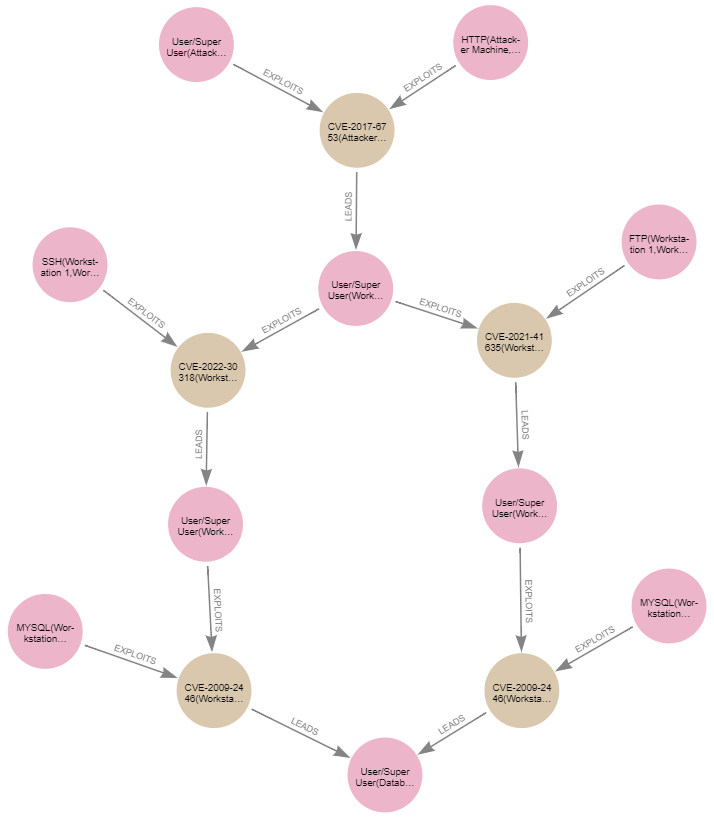}
  \caption{Clinic attack graph.}\label{clinic attack graph}
\end{figure}

\begin{figure}
  \begin{center}
  \includegraphics[width=1.6in]{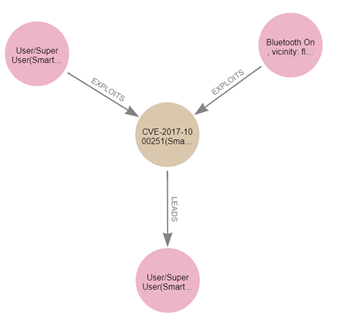}
  \caption{Patient attack graph.}\label{patient attack graph}
  \end{center}
\end{figure}

As mentioned in Section \ref{rw}, we adopt the exploit dependency graph representation \cite{barik2016attack}, and our approach in generating the attack graph for a specific network topology follows similar steps discussed in \cite{barik2016network}. 
\\
However, in addition to all information considered in \cite{barik2016network} to generate an attack graph, in our work we consider the information associated with vulnerabilities such as pre and post conditions which are missing from the work in \cite{barik2016network} which only considers attacker gained privileges. Besides, we explicitly model the accessibility information between hosts in the attack graph generation. 
\\
Moreover, as opposed to the work in \cite{barik2016network} which creates an attack graph merged with topology graph in one graph, our approach needs to read information from the reachability graph and generates an attack graph separate from the topology and the reachability graphs, which makes it easier to query and analyse. Our attack graph can be defined as $G_A= (V, E)$, where $V$ is the set of nodes in the attack graph, and $E$ is the set of edges that connect the nodes in $G_A$. The attack graph involves two types of nodes, such that $V=\{V_c, V_e\}$, where: $V_c=\{V_{pre}, V_{post}\}$, is a \textbf{set} of condition nodes, that can either be pre-conditions required to exploit device or post-conditions resulting from a successful exploit, whereas $V_e$ is a \textbf{set} of vulnerability exploit nodes, that represents an exploit of a vulnerability on a software on a device. 
\\
The attack graphs for the clinic and the patient are shown in Fig.~\ref{clinic attack graph} and Fig.~\ref{patient attack graph} respectively. The pink nodes represent the conditions (either pre- or post-conditions), whilst the brown nodes represent vulnerability exploits.
\\
Generally, in our attack graph we say  $\forall v_v \in V_v$ $\exists v_{pr} \in V_{pre}$ AND $\exists v_{po} \in V_{post}$. Nodes representing pre- and post- conditions will be automatically generated when generating an attack graph. We assume that in order to exploit a vulnerability, all the preconditions have to be met. When it comes to vulnerability exploit nodes, any $v_e \in V_e$ has a name, $v_e.name = CVE-ID(v_{e_i}, v_{e_j})$, where $CVE-ID$ is the unique identifier of the software vulnerability and defined in the $CVE$ database  that can be searched using \cite{21841}. 
\\
$v_{e_i}$ and $v_{e_j}$ are the vertices that represent the end devices in the network topology graph from where the attacker launch the exploit, and the device where the software vulnerability exists respectively. On the other hand, $E=\{E_e, E_l\}$, i.e. in our attack graph, each edge $e \in E$ belongs to one of two sets: $E_e$ and $E_l$, where $E_e$ is a set of edges of category/type EXPLOIT that connect the pre-condition nodes $V_{pre}$ and the nodes $V_e$. $E_l$ is a set of edges of category/type 'LEADS' that connect the exploit  nodes $V_e$ and the achieved post-condition nodes $V_{post}$.
\\
For example, to exploit the  vulnerability \textit{CVE-2017-8628} in the internet browser of Workstation 1, the attacker machine needs to use HTTP protocol to connect to  Workstation 1, which is a pre-condition required to successfully exploit \textit{CVE-2017-8628}. If the attack is conducted successfully, then the attacker will become a user on Workstation 1 as this is the post-condition of the vulnerability. 
\\
Hence, HTTP should be included in the attacker machine accessibility list, as well as in the pre-conditions list associated with \textit{CVE-2017-8628} vulnerability. Having user privileges on  Workstation 1 is one pre-condition required to exploit the FTP vulnerability \textit{CVE-2021-41635} on Workstation 2, which is reachable from Workstation 1. The other pre-condition is that the attacker needs to use FTP protocol to access Workstation 2, hence Workstation 1 needs to support such protocol, i.e., should be included in Workstation 1 accessibility list as well as in the pre-conditions list associated with vulnerability \textit{CVE-2021-41635}, only after which the vulnerability \textit{CVE-2021-41635} can be exploited and the attacker can acquire user privileges on Workstation 2 and proceed to the next vulnerability exploit.
\\
Hence, $\forall precondition \in v_{v_i}.preConditions$ where $v_{v_i} \in  V_v$ our algorithm (Cypher query) creates a  pre-condition node $v_{pr} \in V_{pre}$ in the attack graph $G_A$. Note that, part of the pre-conditions required to successfully exploit a vulnerability is being a user or a superuser on a device from where the attack is launched, hence a pre-condition node representing the privilege required will be created. Similarly, $\forall postcondition \in v_{v_i}.postConditions$ where $v_{v_i} \in  V_v$ our algorithm (cypher query) creates a post-condition node $v_{po} \in V_{post}$ in the attack graph $G_A$. In our use case scenario, we only assume one post-condition  as an outcome from a successful exploit, representing the privilege the attacker gained either as a User or a SuperUser on the attacked device. 
\\
To generate the attack graph, we follow the steps presented in the pseudocode depicted in Listing \ref{code1}:

\begin{lstlisting}[language=python, caption=Attack graph generation, label = code1]
#Step 1:
for(each end device i in reachabilityGraph)
 if(i.privilege='User' OR i.privilege= 'SuperUser')
 
    n1= AttackGraph.add_node(i)
    n1.name = i.privilege
    n1.category = Condition 

#Step 2:
  for(each end device j in reachabilityGraph)
    if(i REACHES j)
      for(each vul v in reachabilityGraph)
        if(j HAS v)
            
#Step 3:
          for(each item x in v.preConditions)  
            for(each item y in j.accessibility)
              if(x == y)
      
#Step 3-a:
                  n1= AttackGraph.add_node(x) #precondition node
                  n1.name = x
                  n1.category = Condition 
           
#Step 3-b:
                  n2= AttackGraph.add_node(v, i, j) #exploit node
                  n2.name =v.name+'('+i+','+ j+')'
                  n2.category = Exploit 
            
#Step 3-c:
                  for(each z in v.postConditions)  
                    n3=AttackGraph.add_node(z) #postconsition node
                    n3.name = z
                    n3.category = Condition 
            
#Step 3-d:
                  e1= attackGraph.add_edge(n1, n2); #exploits edge
                  e1.category = EXPLOITS
           
#Step 3-e: 
                  e2= attackGraph.add_edge(n2, n3); #leads edge
                  e2.Category = LEADS
\end{lstlisting}

The following discusses each of the steps included in pseudo-code presented above.
\begin{enumerate}
    \item For each device where the attacker has acquired User/Super user privileges, create a condition node representing the privilege the attacker has gained on that device.
     \item For each device reachable from the attacker device, and which has one or more vulnerabilities:
      \item  Check whether the pre-conditions to exploit each of those vulnerabilities are satisfied, i.e. check the accessibility list of the device where the attacker acquired User/Superuser privileges (i.e the list of protocols according to which a device can initiate connections), and the pre-conditions list of the vulnerability, if there are accessibility features that match the required pre-conditions to exploit that vulnerability, then:
   \begin{enumerate}
       \item Create a condition node representing each of the satisfied pre-conditions, i.e., matches one or more of the attacker device accessibility features.
       \item Create an exploit node representing the exploit of that vulnerability.
       \item Create a condition node representing each item in the list of post-conditions associated with the vulnerability, i.e., representing the outcome of the successful attack on the device with a vulnerability.
       \item Create EXPOLITS edge between the pre-conditoion node and the exploit node.
       \item Create LEADS edge between the exploit node and the post-condition node.
   \end{enumerate}
\end{enumerate}

\subsubsection{Neo4j Implementation of the Attack Graph $G_A$} \label{NAG}
Using Neo4j, we have implemented Cypher queries that allow us to generate the attack graph from the reachability graph. Following the steps in the pseudocode in Listing \ref{code1}, we can generate the attack graph for a network with a topology graph $G_N$ using Cypher queries. For example, for the clinic topology graph $G_{N_C}$ we implemented the following Cypher queries: 
\begin{Verbatim}[showspaces=false,fontsize=\small]

Step 1: 
MATCH(n:ClinicTopology:EndDevice) 
WHERE n.privilege = ‘User('+n.name+‘)' OR 
n.privilege = ‘SuperUser('+n.name+‘)'  
MERGE(n1:Condition{
name:‘User/SuperUser('+n.name+‘)'})  
ON CREATE SET 
n.targetTopology = ClinicTopology, 
n1:ClinicAttackGraph 

Step 2: 
MATCH(m:ClinicTopology:EndDevice) 
MATCH(n)-[:REACHES]->(m)-[:HAS]->
(v:Vulnerability)

Step 3:
UNWIND n.accessiblity as y WHERE 
ANY(x IN v.preConditions WHERE x = y)

Step 3-a:
UNWIND v.preConditions AS c 
MERGE(n1:Condition{
name:c+‘('+n.name+‘,'+m.name+‘)'}) 
ON CREATE SET n1:ClinicAttackGraph 

Step 3-b: 
MERGE(n2:Exploit{name: 
v.name+‘('+n.name+‘,'+m.name+‘)'}) 
ON CREATE SET n2:ClinicAttackGraph  

Step 3-c: 
UNWIND v.postConditions AS p 
MERGE(n3:Condition{name:p}) 
ON CREATE SET n3:ClinicAttackGraph 

Step 3-d: 
MERGE(n1)-[:EXPLOITS]->(n2)

Step 3-e: 
MERGE(n2)-[:LEADS]->(n3) 

\end{Verbatim}


As we expected in Section \ref{GM}, the generated clinic attack graph reveals two attack paths that can lead the attacker to become a User/SuperUser on the Database Server. For example, to exploit the CVE-2009-2446 vulnerability on the Database Server from Workstation 2 i.e. CVE-2009-2446(Workstation 2, Database) the attacker needs to be a User/SuperUser on Workstation 2, i.e., User/SuperUser(Workstation 2) and also access the Database Server through MYSQL from Workstation 2, i.e., MYSQL(Workstation 2, Database). The successful exploit of CVE-2009-2446(Workstation 2, Database) leads to the attacker acquiring the User/SuperUser(Database) privilege as a post-condition. 
\\
The two attack paths in the clinic attack graph leading the attacker to become User/SuperUser on the Database Server are:
\begin{Verbatim}[showspaces=false,fontsize=\small]

CVE-2017-6753(Attacker Machine, Workstation 1)
->CVE-2021-41635(Workstation 1, Workstation 2)
->CVE-2009-2446(Workstation 2, Database)
\end{Verbatim}
\begin{Verbatim}[showspaces=false,fontsize=\small]

CVE-2017-6753(Attacker Machine, Workstation 1)
->CVE-2022-30318(Workstation 1, Workstation 3)
->CVE-2009-2446(Workstation 3, Database)

\end{Verbatim}
The generated patient attack graph reveals one attack path leading the attacker to become a User/SuperUser on the patient Smart Watch, that is by exploiting the vulnerability CVE-2017-1000251 from the Smart Phone, which leads the attacker to become a User/SuperUser on the Smart Watch, i.e:
\begin{Verbatim}[showspaces=false,fontsize=\small]

CVE-2017-1000251(Smart Phone, Smart Watch)

\end{Verbatim}

\subsubsection{Attack Graph and Cycles}
If the attacker is User/Superuser on a device with accessibility features that allow it open connections and satisfy the pre-conditions required to successfully exploit a software vulnerability exits in a reachable device, allowing the attacker to eventually become User/Superuser on the reachable device, then cycles may occur if the attacker device has also a vulnerability that requires pre-conditions to be exploited from the reachable device if that reachable device has accessibility features that allow it to open connection with the attacker device.  
\\
In other words, cycles in the attack graph may occur if the following condition applies: 
If there is an edge $(v_{e_i}, v_{e_j}) \in E_r$ in the reachability graph ($G_R$), and if the attacker is a User/Superuser on the end device $v_{e_i} \in V_e$ which has an accessibility list $a_{e_i} \in A_v$ and has a software vulnerability $v_{v_i} \in V_v$ that has pre-conditions list $p_{r_i}$ to be successfully exploited, and the end device $v_{e_j} \in V_e$ has an accessibility list $a_{e_j} \in A_v$ and has a software a vulnerability $v_{v_j} \in V_v$ that has pre-conditions list $p_{r_j}$  to be successfully exploited, then if the attacker becomes a User/Superuser (depending on the post-conditions) on the end device $v_{e_j}$ and if the features in the accessibility list $a_{e_j}$ match a set or a subset of the pre-conditions list $p_{r_i}$ required to exploit $v_{v_i}$ exists in the end device $v_{e_i} \in V_e${, i.e., \textit{if} $v_{e_j}.a_{e_j} \subseteq v_{v_i}.p_{r_i}$ then a cycle will be generated. 
\\
To sum up, cycles will appear in the attack graph in this case:
If $(v_{e_i}, v_{e_j}) \in E_r$  and the attacker becomes a User/Superuser on $v_{e_j}$ and there exists $v_{e_j}.a_{e_j}$ such that $v_{e_j}.a_{e_j} \subseteq v_{v_i}.p_{r_i}$. 
\\
If cycles are inevitable in the attack graph and in case one wants to eliminate them, they can do so by first detecting them then remove them. We can do that by finding the shortest path between any two nodes considering the shortest path refers to the path with the minimum number of edges (relationships) between two nodes \cite{bopche2017graph, phillips1998graph}. This is because in the cycle, the start node and the end node is the same node leading to a shortest path between any two nodes in a graph where cycles are inevitable. 
\\
After finding the cycle path, i.e., the shortest path between any two nodes in the graph, then we just remove the exploit nodes from the detected cycle path and all the edges attached with them. To achieve this we implemented the following Cypher query: 

\begin{Verbatim}[showspaces=false,fontsize=\small]

MATCH(m1:ClinicAttackGraph)-[]->
(m2:ClinicAttackGraph)
cyclePath=shortestPath((m2)-[*]->(m1)) WITH m1
nodes(cyclePath) AS cycle WHERE m1:Exploit   
DETACH DELETE m1

\end{Verbatim}
As a result of running the above query, some condition associated with the removed exploit nodes becomes isolated without edges, we can remove those remaining node (condition nodes) without edges, as having those isolated nodes although is not harmful but useless. We can add the following cypher queries, which finds those nodes with zero edges and delete them:
\begin{Verbatim}[showspaces=false,fontsize=\small]

MATCH(n:ClinicAttackGraph) 
WHERE SIZE((n)--())=0 DELETE (n)

\end{Verbatim}
In our attack graph, a cycle will be generated due to the exploit of the vulnerability of the Bluetooth adapter equipped in the Kiosk launched by an attack initiated from Workstation 1, and the exploit of the vulnerability of the Bluetooth adapter equipped in the Workstation 1 launched by an attack initiated from Kiosk. This is because both the Kiosk and the Workstation 1 are reachable from each others (same Bluetooth vicinity) and the the preconditions required to exploit the Bluetooth vulnerabilities in both of the devices are satisfied by each of them. 
\\
However, as the attack target in the clinic topology is the Database Server, having this path (the cycle) from Workstation 1 to the Kiosk then to Workstation 1 is not useful in the clinic attack graph as this path will not lead the attacker to the target, i.e., the Database Server, so we can remove this cycle from our attack graph as it is explained above.  

\section{Merging Graphs} \label{MergingGraphs}
When a sub-network joins another network they may interconnect with each other as it is explained in Section \ref{intro}, the topologies of both networks will be interconnected and the reachability between the devices in both networks will change, as a result, their associated attack graphs will also be interconnected. Hence, in order to depict how attack propagation changes when two networks join each other and interconnected  we need to do the following major steps:
\begin{enumerate}
    \item Update (merge) the topologies of the involved networks, i.e., create new connection links.
    \item Re-generate the reachability graph of the merged topologies.
    \item Re-generate the attack graph of the merged topologies.
    \end{enumerate}
The following sections describes in details each of these steps.

\subsection{Merging network topology graphs}
In essence, if a device has an IP address, when it joins a new system, it will look for an access point (router device) with which a bidirectional TCP connection can be created (in our use case scenario we only consider TCP traffic). However, even if the device does not have an IP address, it still can create point-to-point links with devices in its vicinity. 
\\
Hence, once a device joins a new system, different connectivity edges maybe created. The outcome will be a merged topology comprised of the original topologies of the networks we have in the environment plus newly created CONNECTS\_TO edges between them. In Neo4j, this can be implemented as follows:
\begin{Verbatim}[showspaces=false,fontsize=\small]

MATCH(m1:EndDevice) WHERE 
ANY(x IN m1.accessibility WHERE x =‘IP')  
MATCH(m2:Router) WHERE 
NOT m1.targetTopology = m2.targetTopology

\end{Verbatim}
In the above query we are looking for all the end devices that have IP addresses, and all the routers we have in the environment. We are not filtering a specific network topology (through specifying a category), this is to ensure to look both ways between the two topology graphs in question, i.e., the clinic and the patient as the patient may have routers too. If those devices do not belong to the same network topology, then do the following: 
\begin{Verbatim}[showspaces=false,fontsize=\small]

MERGE(m1)-[:NEW_CONNECTS_TO {via:‘TCP'}]->(m2)
MERGE(m2)-[:NEW_CONNECTS_TO {via:‘TCP'}]->(m1)
SET m1:MergedTopologies, m2:MergedTopologies 

\end{Verbatim}
In the above query, we are connecting the devices from the two network topologies, but with new connection TCP links, i.e., the edges of the new category/type of NEW\_CONNECTS\_TO. As we will discuss in the following Section, this is to distinguish the new created links from the old links, which will help in finding the reachability only for the updated parts of the merged topologies, as the above query allows us find reachability for devices associated with new connections only and avoid re-calculating the reachability for the whole network. 
\\
Thereafter, we add the devices associated with the newly created edges to a new category called \textit{MergedTopologies} which will help us merge and de-merge the associated attack graphs, when the patient leaves the clinic or moves to a different floor. Besides creating new connection edges via TCP, we are creating direct connection links due to the short comm protocols like Bluetooth, such type of connection requires another condition to be satisfied, e.g., the devices has to be in the Bluetooth vicinity of each other. To achieve this we write the following Cypher query:
\begin{Verbatim}[showspaces=false,fontsize=\small]

MATCH(n1:PatientTopology) WHERE 
ANY(x IN n1.accessiblity WHERE x=‘Bluetooth')  
MATCH(n2:ClinicTopology) WHERE 
ANY(x IN n2.accessiblity WHERE x=‘Bluetooth') 
AND n2.floor = $floor

\end{Verbatim}
In the above query, we set the vicinity condition through the property floor, but we allow the user to specify the value of this property at run time through the parameter \$floor to avoid the need to change the value of the floor manually each time the patient moves to a different floor, this is just one way on specifying the vicinity. 
\\
In the above query, we consider only one example of short range communication, that is Bluetooth, but this code can easily be upgraded for all possible short communication technologies that allow different networks merge with each other considering any other condition required to create the connection edge. Next we connect the Bluetooth enabled devices from both networks using the following cypher query:  
\begin{Verbatim}[showspaces=false,fontsize=\small]

MERGE(n1)-[:NEW_CONNECTS_TO{via:‘Bluetooth'}]
->(n2)
MERGE(n2)-[:NEW_CONNECTS_TO{via:‘Bluetooth'}]
->(n1) with n1, n2
n1:MergedTopologies, n2:MergedTopologies} 

\end{Verbatim}
In fact, we can create new connections associated with different typologies that belong to other categories to merge with other networks. For example, imagine if another patient network topology with category PatientTopology2 joins the clinic network, then as an example, we can create new connections associated with the second patient, as follows: 
\begin{Verbatim}[showspaces=false,fontsize=\small]

MERGE(m1)-[:NEW2_CONNECTS_TO {via:‘TCP'}]->(m2)
MERGE(m2)-[:NEW2_CONNECTS_TO {via:‘TCP'}]->(m1)
SET m1:MergedTopologies, m2:MergedTopologies 

\end{Verbatim}
We can do the same for multiple networks joining other networks, each time we distinguish the new connections associated with the new merge using different edge category/type. This guarantees always updating the reachability for only the updated parts of the merged networks and avoid repeating querying all the network to update the reachability results, as it will be explained in details in the following Section.

\subsection{Updating the reachability graph} \label{URG}
Due to the newly created NEW\_CONNECTS\_TO edges in the merged network topology graphs when a network joins another one, the reachability graph of both betworks will inherently change; e.g., new devices such as the patient Smart Phone are merged in the clinic topology and can reach other devices. \\
Thus, we need to re-generate an updated version of the reachability graph by running the following Cypher query, which although looks similar to the reachability query we discussed in Section \ref{RG} that calculates the reachability for whole network, the following query finds the reachability between end devices associated with the newly created connection links in the merged topologies. 
\\
Note that although the reachability query we discussed in Section \ref{RG} can find the reachability in the merged topologies, but this will repeat the work we have already done for the devices in their original network, which already have a poor performance and takes significant time and database hits due to the need to traverse the graph for all indirect TCP connections. 
\\
Hence, we need a reachability query that can find reachability for only the updated parts of the networks in the merged topologies.
\begin{Verbatim}[showspaces=false,fontsize=\small]

MATCH(n:EndDevice)  
MATCH(m:EndDevice)
WHERE n.name <> m.name AND (n.subnet=m.subnet  
OR EXISTS((n)-[:NEW_CONNECTS_TO]->(m))  
OR EXISTS((n)-[:NEW_CONNECTS_TO*{via:‘TCP'}]
->(m)) AND EXISTS((:Router)-[:ALLOWS]->
(:Firewall{source:n.name, destination:m.name}))) 
MERGE(n)-[:REACHES]->(m)

\end{Verbatim}
Note that the newly joined devices may or may not reach existing devices, depending on their connection modes and the firewall rules deployed, as explained in Section \ref{NT}. 
\\
Besides checking the existence of the newly created links in the merged topologies, the above query still check whether end devices belong to the same subnets, which is useful in case one adds a new device to a specific subnet in the current network, which enables the new added device to reach and become reachable from devices in the same subnet. 
\\
However, as in our merged topology query we consider merging two networks by creating new connection links where necessary rather than adding a specific device to a specific subnet, we can omit the first case, i.e., checking whether devices belong to the same subnet when updating the reachability for the merged topologies. Instead, reachability between end devices in the same subnet can be assumed by default whenever one add a new device to a specific subnet.  \\
The following query finds the reachability when a second patient joins the clinic network:
\begin{Verbatim}[showspaces=false,fontsize=\small]

MATCH(n:EndDevice)  
MATCH(m:EndDevice)
WHERE n.name <> m.name AND(
EXISTS((n)-[:NEW2_CONNECTS_TO]->(m)) OR 
EXISTS((n)-[:NEW2_CONNECTS_TO*{via:‘TCP'}]->(m)) 
AND EXISTS((:Router)-[:ALLOWS]->(:Firewall 
{source:n.name, destination:m.name}))) 
MERGE(n)-[:REACHES]->(m)

\end{Verbatim}
We tested our work for multiple patients joining the clinic, and our queries revealed correct results, but due to space limitations we are considering keeping the wok for multiple network joining and leaving each other to another paper submission.

\subsection{Merging attack graphs} \label{MAG}
Due to the updates in the reachability graph, the attack graphs of both the clinic and the patient need to be re-generated. Updating the attack graphs to represent the changes in the topology and the reachability graph requires re-running the query for generating the attack graph presented in Section \ref{NAG} with two modifications: Firstly, any new generated node will be set to belong to a new category that is the MergedAttackGraph (we can check that automatically using the MERGE clause)\footnote{In generating an attack graph, associated nodes will be created from scratch, but when merging attack graphs, some nodes will be already there and new ones will be created, and we don not want to create nodes that are already there}. This is because during the merge of the attack graphs, new edges will be generated in the target attack graph, but new nodes will also be generated. For example, in the clinic attack graph presented in Section \ref{NAG}, we do not have any condition and exploit nodes associated with vulnerabilities exist in the Kiosk software services, this is because the attack graph of the clinic is generated for attacks originating from the Internet and involving attacking the Kiosk in the external attack would result in a cycle, because the attacker would go to the kiosk from the Workstation 1 and back to Workstation 1 to proceed further to the database. Secondly, we need to specify the target topology we are interested in this work each network topology graph in the MergedTopologies category has its own target(s) and thus its own attack graph. For this purpose we use a parameter (using \$) allowing the user to specify the target they are interested in at run time instead of specifying statically as hard-coded value. For example, Step 2 and Step 3-a in Section \ref{NAG} would have the following minor modification: \\
\begin{Verbatim}[showspaces=false,fontsize=\small]
Step 2:
MATCH(m:MergedTopologies:EndDevice) WHERE 
m.targetTopology=$targetTopology 
MATCH(n)-[:REACHES]->(m)-[:HAS]->
(v:Vulnerability)

Step 3-a:
UNWIND v.preConditions AS c 
MERGE(n1:Condition{
name:c+‘('+n.name+‘,'+m.name+‘)'}) 
ON CREATE SET n1:MergedcAttackGraphs

\end{Verbatim}
If we are interested in checking all the updates in all attack graphs associated with all merged topologies, then we need to update (merge) the attack graphs associated with each network topology one by one in an iterative way but each time one need to specify the target they are interested in.

\section{De-merging graphs} \label{DeMergingGraphs}
In brief, when a system (sub-network) leaves, we need to remove its nodes from the merged attack graph and remove all the edges that were added due to the interconnections between the two systems. In short, we need to:
\begin{enumerate}
    \item Update (de-merge) the merged topology graphs by removing the newly added connection links.
    \item Update(de-merge) the merged attack graphs by removing newly added nodes and all edges associated with them.
\end{enumerate}
These steps are explained in the following subsections.

\subsection{De-merging Network Topology Graphs}
As discussed in Section \ref{MAG}, in the merging process, new connection edges are added between the devices from the merged network topology graphs. REACHES edges are also created between devices in the merged topologies due to updating reachability for the updated part of the network. 
\\
Hence, in the de-merging process, we only need to delete the new edges created between the merged topologies without removing the graph nodes representing the devices in the merged networks, in our use case the patient and the clinic networks. Moreover, we need to remove those network components in the merged topology graphs from the MergedTopologies category. 
\\
In other words, we can say: 
$\forall (v_{e_i}, v_{e_j}) \in {E_d}$ and  $\forall e(v_{e_i}, v_{e_j}) \in {E_r}$, where $v_{e_i} \in G_{N_P}$ and  $v_{e_j} \in G_{N_C}$  and both $v_{e_i}$ and $v_{e_j} \in  MergedTopologies$ DELETE $(v_{e_i}, v_{e_j})$ and remove the MergedTopologies from the labels associated with $(v_{e_i}, v_{e_j})$, where $G_{N_P}$ and $G_{N_C}$ refer to the categories of patient and the clinic network topology graphs respectively. The following is the Cypher query for de-merging the merged network topology graphs, i.e., in our use-case scenario the patient and the clinic.
\begin{Verbatim}[showspaces=false,fontsize=\small]

MATCH(n1:PatientTopology)-[r]-(n2:
ClinicTopology) REMOVE 
n1:MergedTopologies, n2:MergedTopologies
DELETE r 

\end{Verbatim}
The above query guarantees removing all the edges between end devices in the MergedTopologies category whether they are NEW\_CONNECTS\_TO or REACHES edges. \\
If there are multiple network topology graphs in the MergedTopologies category, we can filter which network topology graph we are interested in removing from the merged topologies. For example, if there are two patients topologies merged with the clinic network topology, and in case patient 2 leaves the clinic while patient 1 stays in the clinic, we can just edit the above query as follows:
\begin{Verbatim}[showspaces=false,fontsize=\small]

MATCH(n1:PatientTopology2)-[r]-(n2:
ClinicTopology) REMOVE 
n1:MergedTopologies, n2:MergedTopologies
DELETE r 

\end{Verbatim}
The above query will remove the NEW2\_CONNECTS\_TO physical links as well as the REACHES edges associated with patient 2 when their topology merged with the clinic topology. This way, we can just specify the category of the devices we are interested in removing, and use filtering and categories to keep track of the nodes and edges added at each Merge operation.
\\
We tested our work for multiple patients joining the clinic, and our  de-merge queries revealed correct results, but as we mentioned prevouisly, due to space limitations we are considering keeping the wok for multiple network joining and leaving each other to another paper submission.

\subsection{De-merging attack graphs}
When the patient leaves the clinic, we need to update the attack graphs by de-merging the attack graphs before merging them again if necessary e.g if the patient moves to a different floor in the clinic. 
\\
In de-merging the attack graphs, all we need to do is to find the nodes in the temporary category MergedAttackGraphs, which contains the newly added nodes that were not originally in the attack graphs, and delete them along with all the edges that connect them with the rest of the original nodes in the graph. This guarantees separating the original attack graphs corresponding to the network topology graphs, and removing newly temporary created nodes and all edges associated with them. 
\\
In other words, we can say:
$\forall v_{m_a} \in MergedAttackGraphs$ AND $\forall e_a \in E_a$, where $v_{m_a}$ is part of an edge $e_a$ then DELETE $v_{m_a}$ and $e_a$.  We can implement this as a Cypher query as follows:
\begin{Verbatim}[showspaces=false,fontsize=\small]

MATCH(n:MergedAttackGraphs) DETACH DELETE n

\end{Verbatim}
The above query guarantees de-merging all the merged attack graphs. If there is another merged attack graph in the MergedAttackGraph category associated with patient 2 topology graph in the MergedTopologies category, and if patient 2 leaves, we need to specify that in the MergedAttackGraphs category, as follows:
\begin{Verbatim}[showspaces=false,fontsize=\small]

MATCH(n:PatientToplogy2:MergedAttackGraphs) 
DETACH DELETE n

\end{Verbatim}
The above query guarantees only de-merging attack graph of patient 2 from all other merged attack graphs. Thanks to using categories (labels) to filter our queries so to work only on topology and attack graphs we are interested in. 

\section{Running the Algorithms (Cypher queries)} \label{runningAlgo}
To  check the correctness of our algorithms and test whether they generate what is expected, in the following sections we run the queries and show the output for merging and de-merging the graphs associated with the clinic-patient use-case discussed in Section \ref{usecase} on different scenarios related to where the patient is located, i.e., when the patient enters the clinic, and moves from floor to floor. 
\\
We first run the experiments by considering that the target topology is the clinic topology and then we run the same experiments but by considering the target topology is the patient topology. 
\\
We are testing this way because we expect that the merged topologies and attack graphs to change mainly due to the change in the range of Bluetooth enabled devices, as this will change the reachability query output whenever the Bluetooth vicinity changes. The Firewall rules presented in Table \ref{table:1} and \ref{table:2} will not cause changes in the reachablity between devices equipped with IP address when the patient enters and moves between the floors. 

\subsection{Patient in floor1, TargetTopology: ClinicTopology} \label{floor1}
The patient just entered the clinic and is now in floor 1, Fig.~\ref{merged topologies in floor1} depicts how the topology graphs of both the clinic and the patient merges in floor 1, while Fig.~\ref{reachability graph in floor1} depicts how the reachability graphs of both networks will be updated according to the merge. The figures show what it is expected, as the patient's Smart Phone and Smart Watch are Bluetooth enabled, they can reach all Bluetooth enabled devices in their vicinity (floor 1), in this case the Clinic Workstation 1 and the Kiosk. The patient's Smart Phone has an IP address and once in the Clinic, it will look for access points in the clinic to connect to, allowing the Smart Phone to create TCP connections with Router 1 and Router 2.
\\
However, no firewall rule will allow the Smart Phone from reaching clinic end devices using TCP connections in any floor, hence the Smart Phone will only reach clinic Bluetooth enabled end devices in floor 1 through Bluetooth short range communication protocol. On the other hand, although Workstation 1 has an IP addresses, it cannot create TCP connections with any of the patient end devices because the patient does not carry or wear a router device. 
\\
As a result, an internal attacker can exploit the Bluetooth vulnerabilities on both the Workstation 1 and the Kiosk from either the Smart Phone or the Smart Watch, because they satisfy the preconditions required to exploit those vulnerabilities on the clinic end devices located in floor 1 (including the obtaining the required privileges), which is the assumed Bluetooth communication vicinity. The resulted merged attack graphs are shown in Fig.~\ref{merged attack graphs in floor1}. 
\\
Running the reachability query for the updated part of the network as it is explained in Section \ref{URG} took 531 ms to run and required a total of  1681 database hits. This is a significant improvement especially in terms of the total number of database hits as opposed to running the reachability query for the whole network, which will repeat calculating what have been already done, and as our experiment in Section \ref{RG} revealed, running the reachability query for all the end devices in the network required a total of 201529 database hits.
\\
In contrast, the query for merging the topology graphs only took 11 ms and required 616 database hits, while merging the attack graphs took 72 ms and required 4681 database hits. As expected, the de-merge operations are very fast; they  took 3 ms and 4 ms, and required 93 and 38 database hits respectively. As expected, the reachability query is the query that requires the longest time and the most database hits to be processed.

\begin{figure}
  \includegraphics[width=\linewidth]{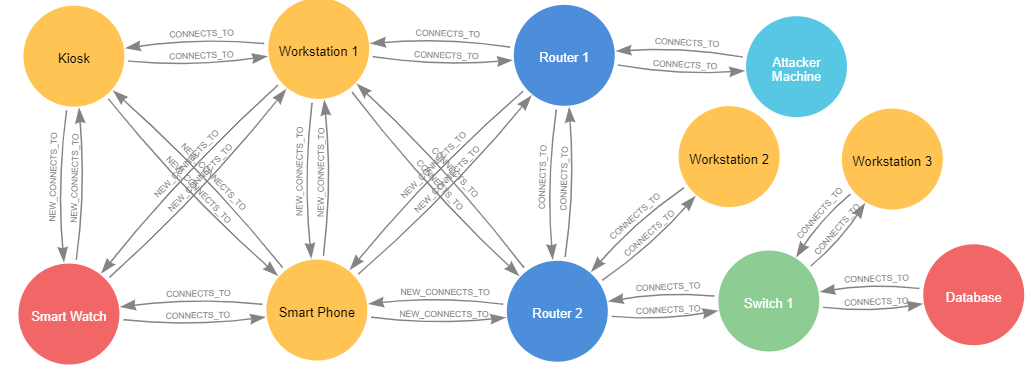}
  \caption{Merged topology graphs in clinic-floor 1.}\label{merged topologies in floor1}
\end{figure}

\begin{figure}
  \begin{center}
  \includegraphics[width=1.8in]{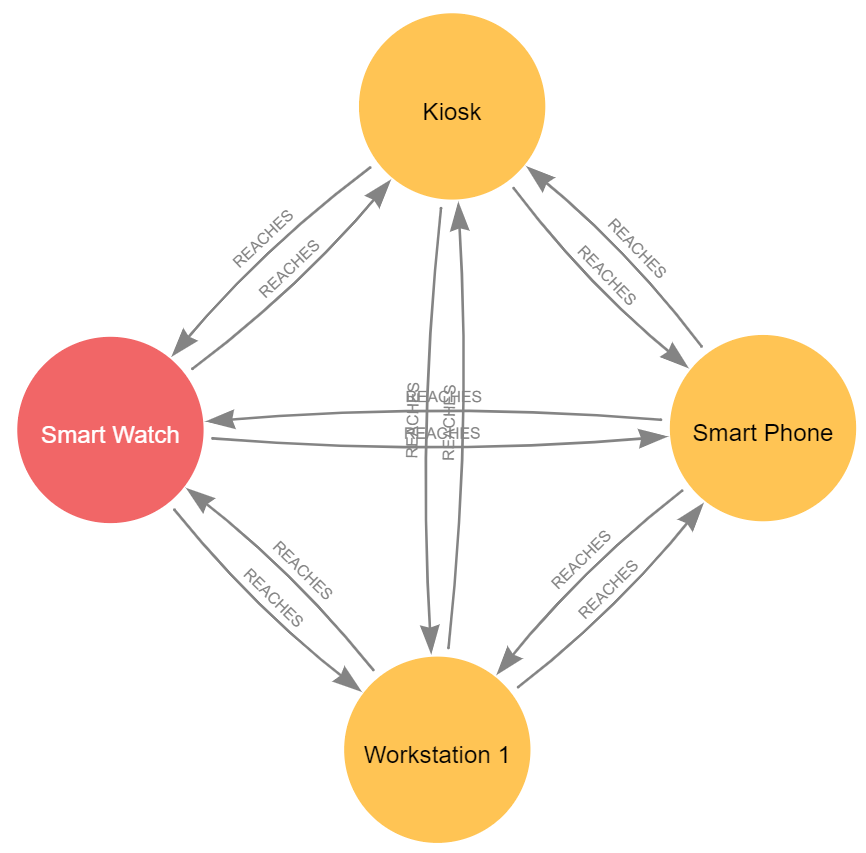}
  \caption{Updated reachability graphs in clinic-floor 1.}\label{reachability graph in floor1}
  \end{center}
\end{figure}

\begin{figure}
  \begin{center}
  \includegraphics[width=\linewidth]{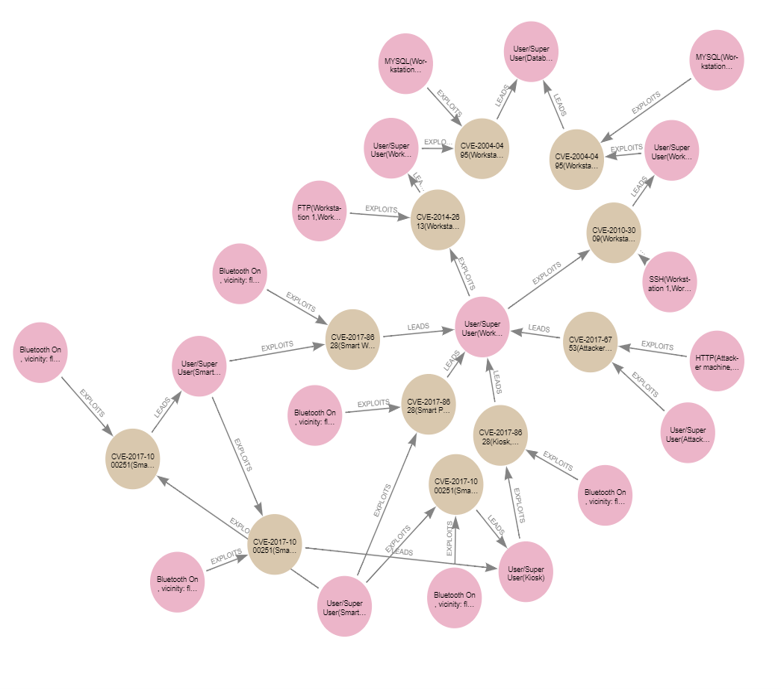}\\
  \caption{Merged attack graphs in clinic-floor 1.}\label{merged attack graphs in floor1}
  \end{center}
\end{figure}

\subsection{Patient in floor2, TargetTopology: ClinicTopology}\label{floor2}
In this section we test our Cypher queries on a different scenario, that is the the patient left floor 1, and moved  to floor 2. Fig.~\ref{merged topologies in floor2} shows how the topology graphs of both the clinic and the patient merges in floor 2, while Fig.~\ref{reachability graph in floor2} shows how the reachability graphs of both networks will be updated according to the merge.
\\
The figures show what is expected, as the patient's Smart Phone and Smart Watch are Bluetooth enabled, they can reach all Bluetooth enabled devices in their vicinity, i.e., floor 2, in this case the Clinic Workstation 2. As patient's Smart Phone has an IP address and once in the Clinic, it will look for access points in the clinic to connect to, allowing the Smart Phone to create TCP connections with Router 1 and Router 2. 
\\
However, no firewall rule allows the Smart Phone from reaching clinic end devices using TCP connections in any floor, hence the Smart Phone will only reach clinic end devices in floor 2 through Bluetooth short range communication protocol. On the other hand, although Workstation 2 has an IP addresses, it cannot create TCP connections with any of the patient end devices because the patient does not carry or wear a router. 
\\
An internal attacker can exploit the Bluetooth vulnerability on Workstation 2 from either the Smart Phone or the Smart Watch, because they satisfy the preconditions required to exploit those vulnerabilities on the clinic Bluetooth enabled  devices located in floor 2, which is the assumed Bluetooth communication vicinity. The merged attack graphs are shown in Fig.~\ref{merged attack graphs in floor2}. 
\\
The reachability query after the merge operation took 222 ms with a total of 1626 database hits (significant improvement as opposed to running the query for the whole network). Merging the topology graphs only took 13 ms and required 402 database hits, while merging the attack graphs query took 11 ms and required 1477 database hits. The queries for de-merging the topologies and the attack graphs took 2 ms and 3 ms, and required 68 and 18 database hits respectively.

\begin{figure}
  \begin{center}
  \includegraphics[width=\linewidth]{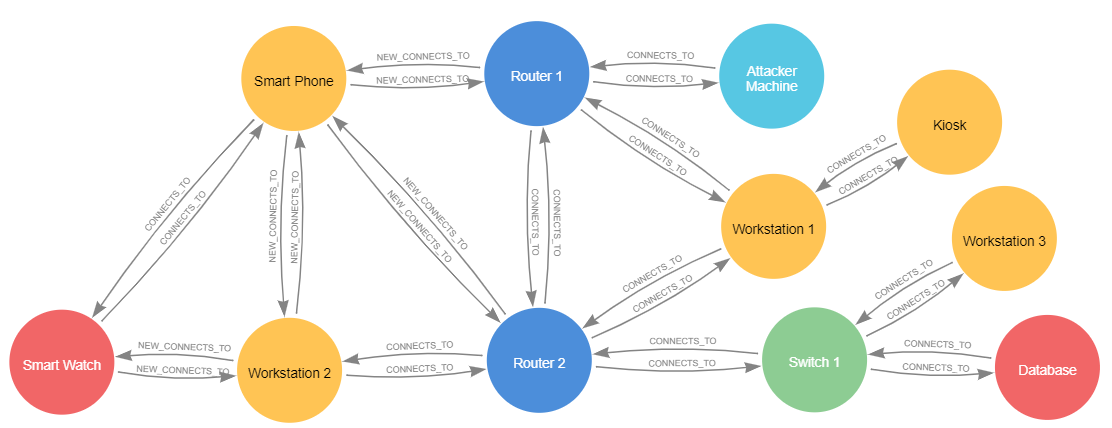}
  \caption{Merged topology graphs in clinic-floor 2.}\label{merged topologies in floor2}
 \end{center}
\end{figure}

\begin{figure}
  \begin{center}
  \includegraphics[width=1.26in]{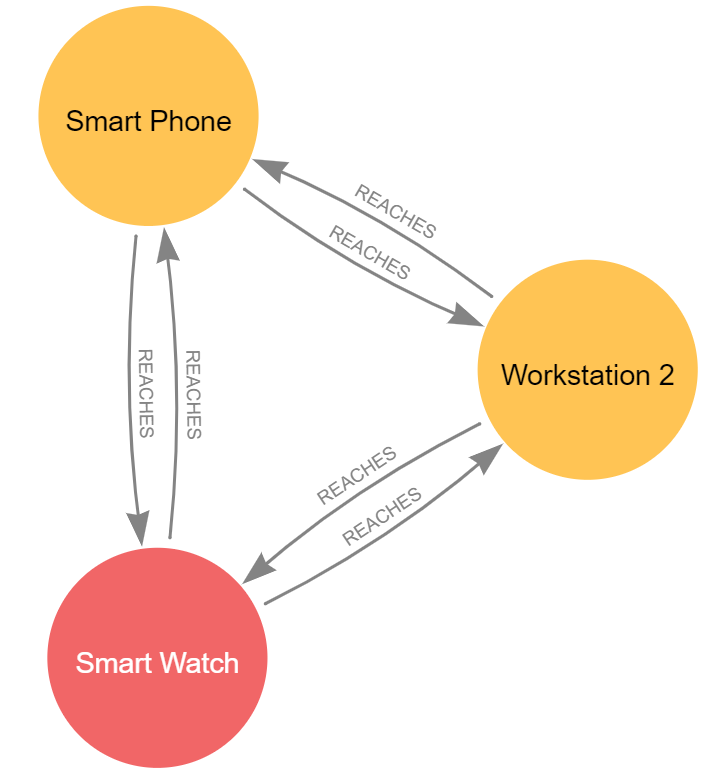}
  \caption{Updated reachability graphs in clinic-floor 2.}\label{reachability graph in floor2}
  \end{center}
\end{figure}

\begin{figure}
  \includegraphics[width=\linewidth]{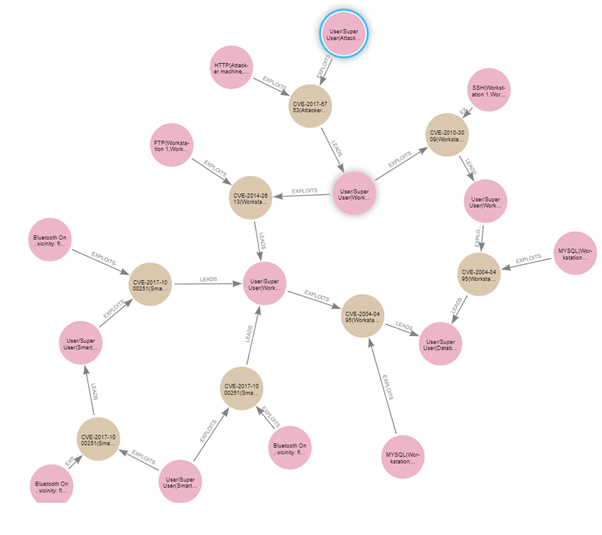}
  \caption{Merged attack graphs in clinic-floor 2.}\label{merged attack graphs in floor2}
\end{figure}

\subsection{Patient in floor3, TargetTopology: ClinicTopology} \label{floor3}
By running this scenario, we show that even if the topologies of different networks interconnect, this does not necessarily mean their associated attack graphs will change, i.e., merged, as the merge of the attack graphs depends on the updates on the reachability graph, if the reachability graphs do not change then the associated attack graphs will not change either. \\
Fig.~\ref{merged topologies in floor3} shows how the topology graphs of both the clinic and the patient merges if the patient leaves floor 2 and moves to floor 3, while Fig.~\ref{reachability graph in floor3} depicts how the reachability graphs of both networks will be updated according to the merge. 
\\
Although the patient Smart Phone will try to open TCP connections with the clinic routers, the attack graphs of the clinic and the patient will not merge. Indeed, neither Workstation 3 nor the Database Server are Bluetooth enabled, and no firewall rule allows communication between the patient’s Smart Phone and any the other end devices in the clinic. 
\\
Although Workstation 3 has an IP addresses, it cannot create TCP connections with any of the patient end devices because the patient does not carry or wear a router device. Accordingly, both the patient and the clinic reachability and attack graphs remain unchanged. 
\\
The reachability query running time after the merge was 216 ms with total 1529 database hits (significant improvement as opposed to running the query for the whole network). The query of merging the topology graphs only took 128 ms and required 246 database hits, while the query of the de-merge of the topologies took 2 ms and required 42 database hits.

\begin{figure}
  \includegraphics[width=3 in]{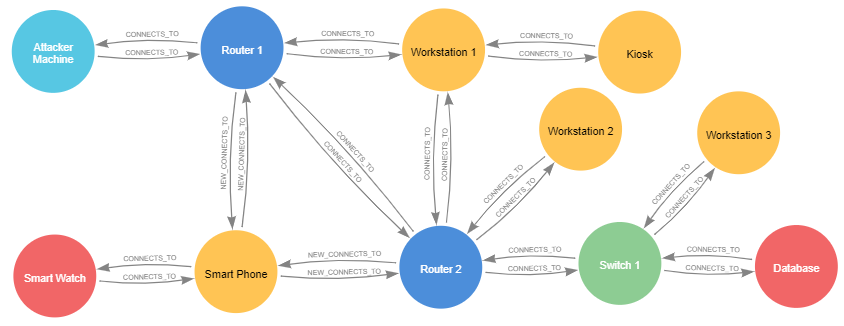}
  \caption{Merged topology graphs in clinic-floor 3.}\label{merged topologies in floor3}
\end{figure}

\begin{figure}
  \includegraphics[width=3in]{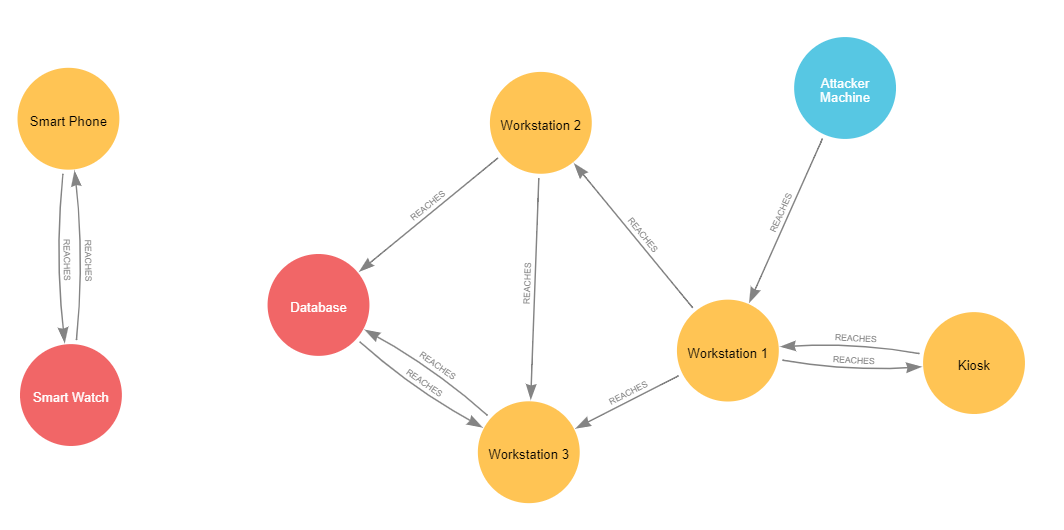}
  \caption{Reachability graphs in clinic-floor 3.}\label{reachability graph in floor3}
\end{figure}


\subsection{Patient in floor1, TargetTopology: PatientTopology}
We continue our queries testing experiments as in Section \ref{floor1}, but we change the target topology to the patient topology. The merged topology and reachability graphs while the patient is in floor 1 of the clinic should be similar to the ones in Fig.~\ref{merged topologies in floor1} and Fig.~\ref{reachability graph in floor1} respectively, which our experimental results revealed. 
\\
The only difference  will be in the merged attack graphs as in this scenario the directions of the directed attack paths are going towards the patient target, i.e., the Smart Watch, with the same number of attack paths (we will discuss how we can count the attack paths in Section \ref{metrics}) leading to the target, i.e., the Smart Watch as what is shown in Fig. \ref{merged attack graphs in floor1-patient}.
\begin{figure}
  \includegraphics[width=\linewidth]{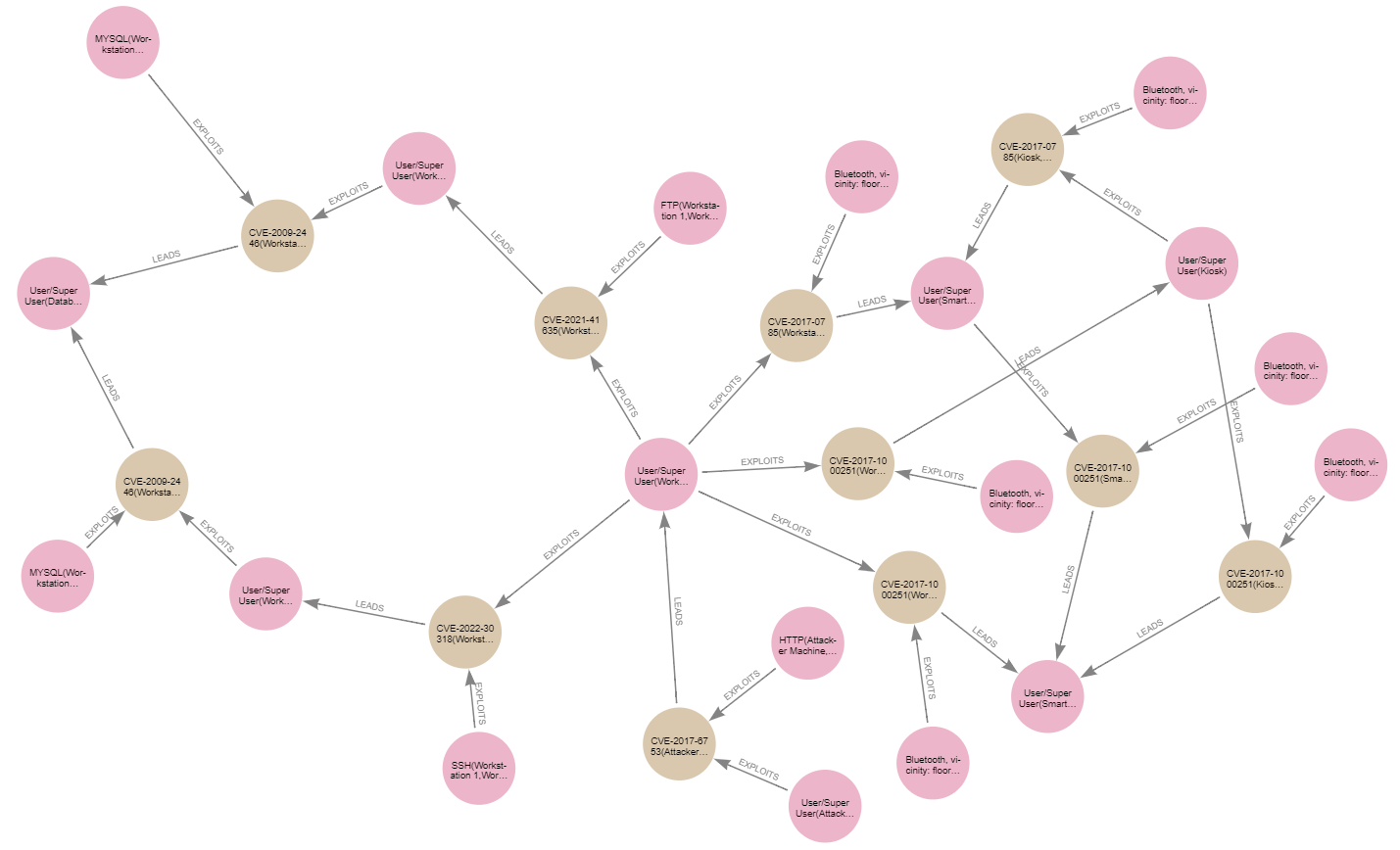}
  \caption{Merged attack graphs in clinic-floor 1, patient as a target topology.}\label{merged attack graphs in floor1-patient}
\end{figure}

\subsection{Patient in floor2, TargetTopology: PatientTopology}
In this Section we run the same experiment as in Section \ref{floor2}, but we change the target topology to the patient topology. The patient left floor 1, and is now in floor 2. As expected, our experimental output revealed that the merged topology and reachability graphs while the patient in floor 2 of the clinic are similar to the ones in Fig.~\ref{merged topologies in floor2} and Fig.~\ref{reachability graph in floor2} respectively. 
\\
The merged attack graphs in floor 2 will be as it is shown in Fig.~\ref{merged attack graphs in floor2}. The directions of the directed attack paths are going towards the patient target, i.e., the Smart Watch, with the same number of attack paths leading to the target as what is shown in Fig. \ref{merged attack graphs in floor1-patient}. This proves the correctness of our queries. 
\begin{figure}
  \begin{center}
  \includegraphics[width=\linewidth]{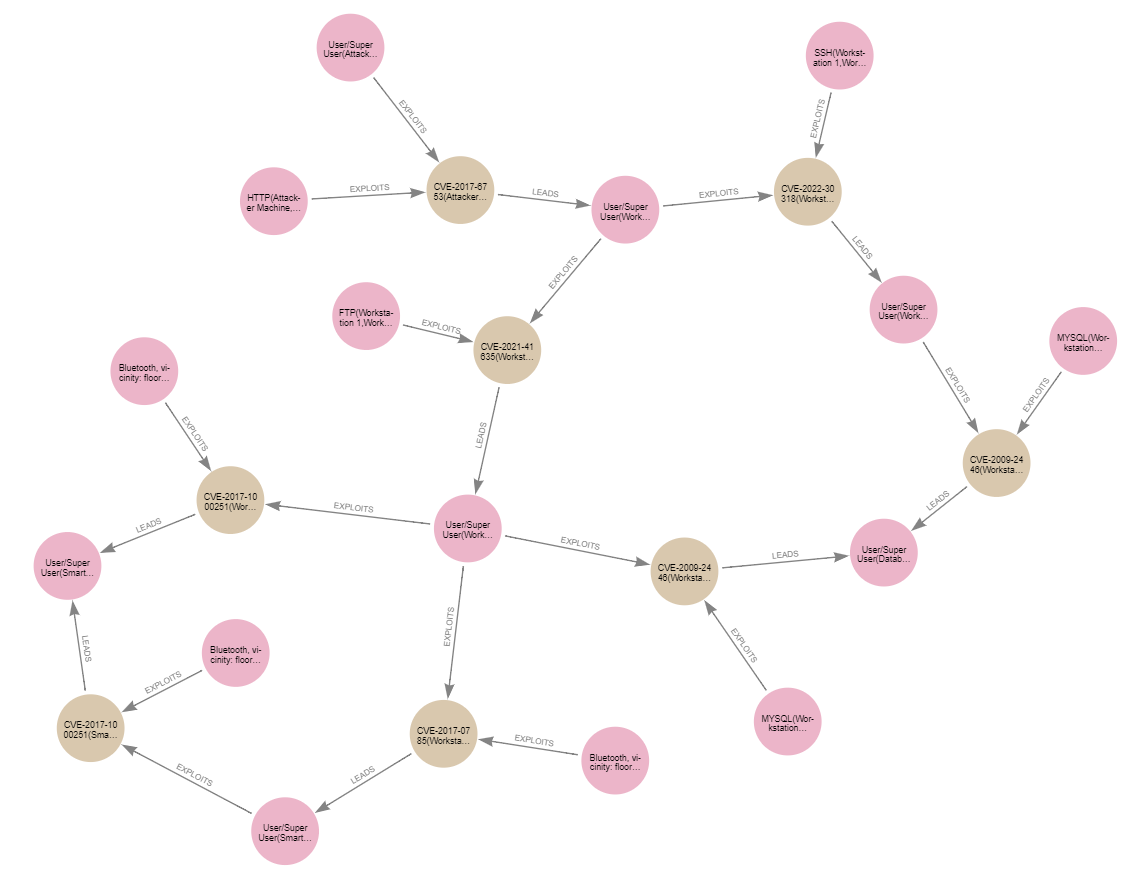}
  \caption{Merged attack graphs in clinic-floor 2, patient as a target topology.}\label{merged attack graphs in floor2-patient}
  \end{center}
\end{figure}

\subsection{Patient in floor3, TargetTopology: PatientTopology}
In this Section we run the same experiment as in Section \ref{floor3}, but we change the target topology to the patient topology. The patient left floor 2, and is now in floor 3. As expected, the merged topology graphs and the updated reachability graphs while the patient in floor 3 of the clinic are similar to the ones in topology  Fig.~\ref{merged topologies in floor3} and Fig.~\ref{reachability graph in floor3} respectively.
\\
Similar to what has been discussed in Section\ref{floor3}, both the patient and the clinic reachability and attack graphs will remain as they were originally, as the patient Smart Phone will try to open a TCP connections with the clinic routers, but the attack graphs of the clinic and the patient will not merge (neither Workstation 3 or the Database Server have Bluetooth enabled, and there is no firewall rule allows communication between the patient’s Smart Phone device and other devices in the clinic. 

\section{Queries on attack graph-based metrics towards risk assessment} \label{metrics}
As we discussed in  the previous sections, the attack graph is a modelling technique used to determine threats to the critical system resources. Determining the possibilities of a cyber attack against critical resources, the potential impact of the attack on other parts of the system, and planing security countermeasures to prevent attacks on a system all fall under the umbrella of the concept of security risk assessment \cite{landoll2021security}. In its core, the attack graph provides a graphical security modelling to help us asses the risk of attack to system components \cite{bopche2017graph}. 
Our proposed method provides a means to assess risk and attack propagation that allows to re-evaluate the risk of compromise to different parts of the system as some parts of the system may have been compromised.
\\
Usually we asses the risk of attacks in a system against specific security metrics in order to identify and quantify security issues early enough so that we select appropriate countermeasures. Security metrics are used to asses the security of a given network configuration and to evaluate day-to-day changes in the security strength of a network \cite{bopche2017graph}. \\
There are a number of metrics available in literature for security risk assessment summarised in \cite{bopche2017graph}, such as the attack path length and non path length based metrics,  which were proposed in \cite{10.1145/310889.310919, ortalo1999experimenting, li2006cluster, idika2010extending, lippmann2006validating, pamula2006weakest}, besides a group of network hardening metrics suit proposed by Noel and Jajodia in \cite{noel2014metrics}, the probability-based metrics proposed in \cite{wang2007measuring, wang2008attack} which are sensitive to the introduction of a new vulnerability in the network, and the graph distance metrics discussed in \cite{bopche2017graph}, which determine changes in the network attack surface at a finer granularity and complements the traditional network monitoring approach for security risk mitigation. \\
In this section, we implement examples on Cypher queries for two attack graph based metrics,  one of which is shortest path metric \cite{10.1145/310889.310919} and the other represents the number of paths metric proposed in \cite{ortalo1999experimenting}.

\subsection{Shortest Attack Path Metric}
The Shortest Path metric represents the length of the shortest distance from an attacker’s initial state to the attacker’s goal state (i.e., the target). In our attack graph the Shortest Attack Path means the path with the minimum number of edges (relationships) from the attacker starting point until the attacker goal node \cite{bopche2017graph, phillips1998graph}. The category/type of nodes that can be included in the path is subject to the security expert performing the analysis. For example, we may want to involve the condition nodes, or we may want to count/ view only the vulnerability exploit nodes along the path or we may want to analyse both \cite{idika2010extending}. In this work we consider the number of conditions and exploits starting from the attacker’s initial state and proceeding in a chain to the attacker’s goal state. But how is this metric useful? \\
Intuitively speaking, from the perspective of the attacker, if there are multiple ways the attacker can follow to get to the target, the attacker will choose the path that requires the least number of steps. This relies on two assumptions: first that the attacker is interested in using the minimum effort to reach the goal node and second, that each step taken by an attacker includes a risk of being detected, so a minimum set of steps would, in principle, also minimise the attacker’s risk of being detected. Hence, the network with the shortest attack path is the network that is less secure, and the security engineer may prioritise patching the software vulnerabilities exist along the shortest attack path.
\\
However, although the shortest path attack metric can indicate the level of the security of a network, this metric is not accurate in all cases. For example, as is discussed in \cite{ortalo1999experimenting}, this metric alone does not indicate how many shortest paths exist in a network, if there more than one exists, which would lead to erroneous analysis and decisions.  Moreover, this metric is not sensitive enough to be used for real-time network security evaluation independently, because if the security of the network is improved by implementing counter measures, this may result in a new attack graph for a more secure network but may have the same minimum path length compared to not implementing any countermeasure. Finally, the assumption that an attacker may choose the shortest path, does not necessarily hold attacker skill e.g., in exploiting a particular type of vulnerability or tooling may induce different attack paths to be chosen. As a result, the Shortest Path metric can be too coarse \cite{idika2010extending}.
\\
Nevertheless, this metric can still be effective to estimate how the security of the network can change for example, when a new vulnerability is discovered, or new countermeasures are being deployed. 
\\
Using the Neo4j Cypher language, we developed queries which allows identifying, visualising and measuring the shortest attack path in a network. This metric can be applied and evaluated when the graph changes e.g., because different systems have been merged e.g., in our running example when the patient is in clinic floor 1. The following query counts the shortest attack path and the length of that path between the patient smart phone and the clinic database server when the patient is in clinic floor 1. This path is depicted in Fig. \ref{shortest attack path in merged graphs in floor1}. We can do the same analysis for each network individually and for any other two nodes, located in the same or different topologies and with multiple targets.
\begin{Verbatim}[showspaces=false,fontsize=\small]

MATCH(start:PatientTopology:EndDevice{name:
'Smart Phone'})
MATCH(end:ClinicTopology:EndDevice{name:
'Database'})
MATCH p = shortestPath((n:PatientAttackGraph:
Condition{name:'User/SuperUser('+start.name+')'
})-[*]->(m:Condition:ClinicAttackGraph
{name:'User/SuperUser('+end.name+')'}))
RETURN  length(p)
OUTPUT:   6
\end{Verbatim}

\subsection{Number of Attack Paths Metric}
The number of attack paths metric represents the number of ways an attacker can compromise a target in a given system and is an estimate of the network exposure to an attack \cite{bopche2017graph, phillips1998graph}. Intuitively, if the attacker has multiple ways to compromise a target node, the attacker has a better chance of achieving their goal without being detected. Hence, an attack graph with the larger number of paths is considered less secure. 
\\
This metric allows to detect fine granular changes in network security that the Shortest Attack Path metric fails to detect. However, it may not alone give an accurate picture about the security of the network, because the attacker effort is not considered in this metric \cite{idika2010extending}.  For example, each path in a graph $G_a$ of 10 attack paths could require 20 times more effort to violate than a graph $G_b$ with one path, which means that $G_b$ is less secure compared to $G_a$  \cite{idika2010extending}. However, there is no known way to quantify the required effort to propagate throughout the attack path in practice. \\
For instance, the following query counts the total number of attack paths between the attacker machine on the internet and the target, i.e., the Database Server in the clinic topology:
\begin{Verbatim}[showspaces=false,fontsize=\small]

MATCH (start:ClinicTopology:EndDevice{name:
‘Attacker Machine'})
MATCH(end:ClinicTopology:EndDevice{name:
‘Database'})
MATCH path =(n:ClinicAttackGraph:Condition{
name:‘User/SuperUser('+start.name+‘)'})-[*]->
(m:ClinicAttackGraph:Condition{name:
‘User/SuperUser('+end.name+‘)'})
RETURN count(path)
Output: 2

\end{Verbatim}
Similar to the above, queries can be implemented to reason about attack paths and metrics in the patient attack graph. We can do the same analysis for each network individually and for any other two nodes, located in the same or different topologies and with multiple targets. For example, the following query counts the number of attack paths between the patient Smart Phone and the target in the clinic, i.e., the Database Server when the patient is in the clinic floor 1, and reveals there are \textbf{8} attack paths between the Smart Phone and the Database Server:
\begin{Verbatim}[showspaces=false,fontsize=\small]

MATCH(start:PatientTopology:EndDevice{name:
‘Smart Phone'})
MATCH(end:EndDevice:ClinicTopology{name:
‘Database'})
MATCH path = (n:PatientAttackGraph:Condition{
name:‘User/SuperUser('+start.name+‘)'})-[*]->
(m:Condition:ClinicAttackGraph{name:
‘User/SuperUser('+end.name+‘)'})
RETURN count(path) 
Output: 8 

\end{Verbatim}
The following Cypher query lists each of those 8 attack paths along with its length:
\begin{Verbatim}[showspaces=false,fontsize=\small]

MATCH(start:PatientTopology:EndDevice{name:
'Smart Phone'})
MATCH(end:ClinicTopology:EndDevice{name:
'Database'})
MATCH path=(n:PatientAttackGraph:Condition{
name:'User/SuperUser('+start.name+')'})-[*]->
(m:Condition:ClinicAttackGraph
{name:'User/SuperUser('+end.name+')'})
RETURN  count(p), length(p)
OUTPUT:
count(p)    length(p)
  2             6
  4             8
  2             10
\end{Verbatim}
By running same query to count the total number of attack paths between the patient Smart Phone and the clinic Database Server (the target) while the patient is in clinic floor 2, we found out that we have \textbf{2} attack paths. Fig. \ref{shortest attack path in merged graphs in floor2} shows the shortest path between the patient Smart Phone and the database while the patient is in clinic floor 2. The shortest path is of \textbf{4} edges as opposed to the other attack path which are of \textbf{6} edges.
When the patient is in the clinic floor 3, running the total number of attack paths query results in an output of \textbf{0} attack paths for the reasons explained in Section \ref{runningAlgo}.
\\
We can run similar queries to check the total number of attack paths and the shortest attack path between the patient Smart Watch and the Database Server in the clinic. We can also do the same by changing the target topology to the patient topology and checking the total number of paths and the shortest path between any end device in the clinic and the patient target, i.e., the Smart Watch. 
\begin{figure}
  \includegraphics[width=\linewidth]{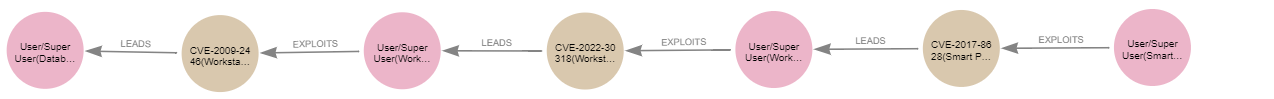}
  \caption{Shortest path query output from the patient Smart Phone and the Database Server in the clinic, while patient is in floor 1.}\label{shortest attack path in merged graphs in floor1}
\end{figure}

\begin{figure}
  \begin{center}
  \includegraphics[width=\linewidth]{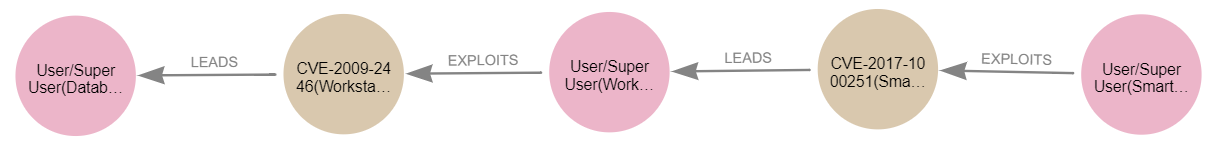}
  \caption{Shortest path query output from the patient Smart Phone and the Database Server in the clinic, while patient is in floor 2.}\label{shortest attack path in merged graphs in floor2}
  \end{center}
\end{figure}

\section{Conclusion} \label{conclusion}
This study has investigated the propagation of attacks in current and future IoT systems. In particular an approach is developed to enable representing and maintaining attack paths through the system whilst allowing for system dynamics, considering not only the addition/removal of single devices but also the merging of graphs when multiple systems are connected and disconnected. This includes considering aspects of system topology and compositionality.
\\
Furthermore, the proposed method provides means to assess risk and attack propagation that allows to re-evaluate the risk of compromise to different parts of the system. \\
We harnessed the efficiency of Neo4j, a popular graph-based tool, to map paths across the network. Neo4j allows us to model network and attack graphs, perform queries that provide quick results, also allowing for the filtering of large graph and query subgraphs of interest, which saves search and query time complexities. For example, when the updates happen dynamically in the networks, the presented reachability query  finds the paths associated only with the newly created links i.e., the updated part of the network, rather than querying all the paths which would have have a poor performance on a scalable network due to the need to traverse the graph for all indirect connections. \\
In comparison with previous reachability query approaches, that query all paths with no consideration for avoiding duplicate calculations, the proposed query is not re-generating everything all the time. We showed the extent to which a graph data-base tool and the optimised implementation of queries on graphs make realising the solution to system dynamics and reachability problem in dynamic graphs possible, by specifying exactly where and how to traverse certain parts of the network: using different edge types, filtering, categories, parameters, attributes and so on. This optimised treatment of graphs allows parts of the system to be queried even  before any updates take place or in static networks, as one does not need to query the whole network for reachability between all network components. \\
The discussed reachability query is the only query that requires traversing a graph searching for indirect connections between end devices and compared to the rest of Cypher queries we proposed that all only require checking directly connected nodes, it takes the longest time to be processed requiring the highest number of database hits, which makes the processing time of the rest of all other queries we proposed in this work negligible. However, after running and testing the proposed queries on different scenarios, results revealed that even when two systems are merged, running the reachability query on the merged graphs provided results in less than a second with the number of database hits reduced significantly as compared to running the query for the whole system. The proposed network topology, reachability and attack graphs models can easily be extended by only creating new nodes, relationships, categories and attributes with no changes required to the database schema. \\
Besides investigating the propagation of attacks in current and future IoT systems, we developed queries to enable us asses the the risk of compromise in the dynamic IoT environment against security metrics evaluated on the graph of the different attack paths. As discussed earlier, the shortest path metric can be too coarse, while the number of attack paths offers a measure of the degree of flexibility the attacker has but does not necessarily capture an appropriate measure of risk to the system. It is therefore crucial to be able to analyse multiple metrics together because as any single metric used in isolation may lead to a misleading analysis and conclusion of the risk to the system \cite{bopche2017graph, phillips1998graph}. 
\\
We therefore have expanded the proposed method to include other metrics for risk assessment to analyse the security strength of networks in a dynamic IoT context, i.e., when two systems join or leave each other, also the method has been enhanced to be able to estimate the impact of compromise upon the functionality of the system. However, due to space limitations, we are not providing this part of the work here,  but rather we are working on another paper to publish the new work. 
\\
Additionally in the short term, we are planning to continue the implementation of the algorithms, evaluating them on scalable and realistic network data-sets. Subsequently, we will be working on developing techniques to select the most appropriate threat mitigation technique to deploy when the IoT environment changes. We aim to achieve this by fully estimating the risk and the impact of compromise when the environment changes, not only by using the methods implemented in this work, but also through Bayesian inference according to the work we have previously done in \cite{munozbayesian}.
\\
The end goal we are seeking is the delivery of implemented tools and algorithms that can be either used independently for the analysis of more dynamic IoT scenarios or integrated into the broader tools of attack graph analysis and risk evaluation developed to model threats in Enterprise networks to enable them to deal with dynamic network scenarios. 

\section*{Acknowledgment}
I thank PETRAS National Centre of Excellence for IoT Systems Cybersecurity for funding this work. \textit{https://petras-iot.org/}

\ifCLASSOPTIONcaptionsoff
  \newpage
\fi

\bibliographystyle{IEEEtran}
\bibliography{IEEEabrv,Bibliography}

\vfill

\end{document}